%% file: beers2014.tex
\documentclass[iop,revtex4]{emulateapj}

\newcommand*{\teff}{$T_{\rm eff}$}
\newcommand*{\logg}{$\log~g$}
\newcommand*{\feh}{[Fe/H]}

\newcommand*{\cfe}{[C/Fe]}
\newcommand*{\zmax}{$Z_{\rm max}$}
\newcommand*{\rapo}{$r_{\rm apo}$}
\newcommand*{\rperi}{$r_{\rm peri}$}
\newcommand*{\vphi}{$V_{\rm \phi}$}

\newcommand*{\rsun}{$R_\odot$}

\newcommand*{\kms}{km~s$^{-1}$}

\newcommand*{\mvphi}{$\langle V_{\phi}\rangle$}

\usepackage{lscape}
\usepackage[dvips]{color}
\usepackage{hyperref}
\usepackage{graphicx}

\hypersetup{
    bookmarks=true,         % show bookmarks bar?
    unicode=false,          % non-Latin characters in Acrobats bookmarks
    pdftoolbar=true,        % show Acrobats toolbar?
    pdfmenubar=true,        % show Acrobats menu?
    pdffitwindow=true,     % window fit to page when opened
    pdfstartview={FitH},    % fits the width of the page to the window
    pdftitle={},    % title
    pdfauthor={Beers},     % author
    pdfsubject={Astronomy},   % subject of the document
    pdfcreator={dvipdf},   % creator of the document
    pdfproducer={dvipdf}, % producer of the document
    pdfkeywords={metal-poor stars},% list of keywords
    pdfnewwindow=true,      % links in new window
    colorlinks=true,       % false: boxed links; true: colored links
    linkcolor=red,          % color of internal links
    citecolor=blue,        % color of links to bibliography
    filecolor=magenta,      % color of file links
    urlcolor=cyan,           % color of external links
    breaklinks=true,
    linktocpage
}

\begin{document}

\title{Population Studies. XIII. A New Analysis of the Bidelman-MacConnell 
``Weak-Metal'' Stars --  Confirmation of Metal-Poor Stars
in the Thick Disk of the Galaxy }

\author{
Timothy C.   Beers\altaffilmark{1,2},
John E.     Norris\altaffilmark{3},
Vinicius M. Placco\altaffilmark{4},
Young Sun      Lee\altaffilmark{5},\\
Silvia       Rossi\altaffilmark{6},
Daniela    Carollo\altaffilmark{7},
Thomas    Masseron\altaffilmark{8}
}

\altaffiltext{1}{National Optical Astronomy Observatory, Tucson, AZ 85719, USA. 
\href{mailto:beers@noao.edu}{beers@noao.edu}}
\altaffiltext{2}{JINA: Joint Institute for Nuclear Astrophysics.}

\altaffiltext{3}{Research School of Astronomy and Astrophysics, The
Australian National University, Mount Stromlo Observatory, Cotter
Road, Weston, ACT 2611, Australia. \href{mailto:jen@mso.anu.edu.au}{jen@mso.anu.edu.au}}

\altaffiltext{4}{Gemini Observatory - Northern Operations Center, 
Hilo, HI 96720, USA. \href{mailto:vplacco@gemini.edu}{vplacco@gemini.edu}}

\altaffiltext{5}{Department of Astronomy, New Mexico State University, Las Cruces, NM 88003, USA. 
\href{mailto:yslee@nmsu.edu}{yslee@nmsu.edu}}

\altaffiltext{6}{Instituto de Astronomia,  Geof\'{i}sica e Ci\^{e}ncias Atmosf\'{e}ricas, 
Departamento de Astronomia, Universidade de S\~{a}o Paulo, Rua do Mat\~{a}o  1226, 
05508-900 S\~{a}o Paulo, Brazil. \href{mailto:rossi@astro.iag.usp.br}{rossi@astro.iag.usp.br}}

\altaffiltext{7}{Department of Physics and Astronomy - Astronomy, Astrophysics and 
Astrophotonic Research Center - Macquarie University - North Ryde, 2019, NSW, Australia.
\href{mailto:daniela.carollo@mq.edu.au}{daniela.carollo@mq.edu.au}}

\altaffiltext{8}{Institute of Astronomy, University of Cambridge, Madingley Road,Cambridge 
CB3 0HA, UK. \href{mailto:tpm40@ast.cam.ac.uk}{tpm40@ast.cam.ac.uk}}

\begin{abstract} 

A new set of very high signal-to-noise ($S/N > 100/1$), medium-resolution ($R
\sim 3000$) optical spectra have been obtained for 302 of the candidate
``weak-metal'' stars selected by Bidelman \& MacConnell. We use these data to
calibrate the recently developed generalization of the SEGUE Stellar Parameter
Pipeline, and obtain estimates of the atmospheric parameters ($T_{\rm eff}$,
log~$g$, and [Fe/H]) for these non-SDSS/SEGUE data; we also obtain estimates of
[C/Fe]. The new abundance measurements are shown to be consistent with
available high-resolution spectroscopic determinations, and represent a
substantial improvement over the accuracies obtained from the previous
photometric estimates reported in Paper I of this series. The apparent offset in
the photometric abundances of the giants in this sample noted by several authors
is confirmed by our new spectroscopy; no such effect is found for the dwarfs.
The presence of a metal-weak thick-disk (MWTD) population is clearly supported
by these new abundance data. Some 25\% of the stars with metallicities $-1.8 <
$ [Fe/H] $\le -0.8$ exhibit orbital eccentricities $e < 0.4$, yet are clearly
separated from members of the inner-halo population with similar metallicities
by their location in a Lindblad energy vs. angular momentum diagram. A
comparison is made with recent results for a similar-size sample of RAVE stars
from Ruchti et al. We conclude, based on both of these samples, that the MWTD is
real, and must be accounted for in discussions of the formation and evolution of
the disk system of the Milky Way.

\end{abstract} 


\section{Introduction}

\subsection{Historical Overview}

In the first paper of this series, \citet[][NBP, Paper I]{norris1985}
presented DDO photometric estimates of metallicity, spectroscopic
measurements of radial velocities, and photometric classifications for a
sample of 309 non-kinematically selected stars taken from the list of
``weak-metal'' candidates originally identified by
\citet{bidelman1973}. Based on these data, and supplemented with
proper motions that were available at the time, NBP obtained space
motions and estimates of orbital eccentricities for a subset of this
sample. Inspection of this distribution led these authors to conclude
that there exists a substantial number of low-metallicity stars on
low-eccentricity orbits ($e < 0.4$), which they found difficult to
reconcile with the classical rapid collapse model for the formation of
the Galaxy put forward by \citet{eggen1962}. NBP suggested instead that
the low-metallicity, low-eccentricity stars belong to a population that
is ``(i) ... intermediate in its motion perpendicular to the Galactic
plane between that of the thin disk and that of metal-deficient objects
of extreme eccentricity, and (ii) that the velocity dispersion of this
group of stars is consistent with its belonging to the thick-disk
population described by \citet{gilmore1984}.'' This suggested population
has become known as the metal-weak thick disk \citep[MWTD -
e.g.][]{morrison1990,beers1995s}.

Attempts to confirm or refute the existence of a MWTD population have
led to numerous (and ever more-detailed) studies over the past two
decades.  \citet{morrison1990} provided additional support for the
MWTD, based on a (slightly revised) DDO photometric abundance scale,
and the kinematics of a low-latitude sample of giants selected to test
for a separation of halo-like and disk-like objects. However, the
conclusions of both of these efforts were called into question by
subsequent work. In the case of NBP, \citet{anthony1994} obtained an
improved calibration of the DDO abundance estimates for their sample
giants, and concluded that there existed an offset of about 0.5 dex
for giants of intermediate metallicity (around [Fe/H]
= $-1.2$). As a result, these authors suggested that the numbers of
stars with disk-like motions and metallicities [Fe/H] $ < -1.0$ in the
work of NBP had been substantially overestimated, compromising the
claim for a MWTD population.  \citet{ryan1995} sought to resolve these
discrepancies by obtaining high-resolution spectroscopic abundance
determinations of some 30 giants in these two samples with claimed
photometric abundance estimates [Fe/H] $< -1.0$. The results of their
study indicated that many, but not all, of the giants in the NBP and
the \citet{morrison1990} sample possessed higher metallicities than
had been inferred from the DDO photometry. As a consequence, they
argued that, although a MWTD may indeed exist, its contribution to the
populations of stars within 1 kpc of the Galactic disk had likely been
overestimated by previous work.

Other observational efforts have addressed the problem of the existence
of a MWTD component in the Galaxy. For instance, \citet{beers1995s}
argued from their sample of non-kinematically selected stars that a
surprisingly large fraction of the metal-poor stars ($ > 30$\% of stars
with [Fe/H] $< -1.5$, rising to 60\% for stars with $-1.6 \le $ [Fe/H] $
\le -1.0$) in the Solar Neighborhood might be associated with a MWTD
component. \citet{chiba1998} used high-quality proper motions from the
Hipparcos satellite for a much smaller sample of red giants and RR
Lyraes to argue that, while a MWTD appeared present, the fractions of
stars at low metallicity associated with it were substantially smaller,
roughly 10\% for stars in the interval $-1.6 \le $ [Fe/H] $ \le -1.0$.
\citet{martin1998} considered the space motions of nearby RR Lyrae stars
with well-determined kinematics, and concluded that a MWTD existed in
their sample (including stars with metallicities as low as [Fe/H] $\sim
-2.0$), similar to previous results for the sample of RR Lyraes examined
by \citet{layden1995}.

\citet{chiba2000} performed a detailed analysis of a large sample of
non-kinematically selected stars with available (medium-resolution)
spectroscopic abundances, radial velocities, and (for roughly half of
their sample) proper motions from the assembly of \citet{beers2000}.
These authors concluded that the fraction of likely MWTD stars in the
Solar Neighborhood with $-1.7 < $ [Fe/H] $\le -1.0$ was on the order of
30\%, falling to on the order of 10\% for stars with $-2.2 < $ [Fe/H] 
$\le -1.7$. \citet{beers2002} analyzed a sample of candidate
low-metallicity giants located close to the Galactic plane from the LSE
survey of \citet{drilling1995}. Their Monte-Carlo experiments on the
distribution of orbital eccentricities of this sample suggested that the
fraction of MWTD stars with [Fe/H] $< -1.0$ might actually be as high as
40\%, and that it may remain as high as 30\% for stars with [Fe/H] $<
-1.6$. Beers et al. reasoned that the origin of this discrepancy with
respect to the work of \citet{chiba2000} came from the selection
criteria employed by most surveys for low-metallicity stars, which
understandably concentrated on regions of the Galaxy with latitudes
above $|b| = 30^\circ$.   

\citet{arifyanto2005} re-analyzed the kinematically selected sample of 
\citet{carney1994}, using Hipparcos-based parallaxes and (where available) 
Hipparcos and Tycho-2 proper motions, and applying corrections to the
Carney et al. photometric distance estimates based on the overlap of the
two samples. Their analysis also indicated the presence of a MWTD
component, with a local stellar fraction smaller than that
claimed by \citet{chiba2000} in the metallicity interval $-1.7 < $ [Fe/H]
$\le -1.0$ (18\% vs. 30\%). They interpreted the origin of the MWTD in
terms of the debris of a ``shredded satellite'', similar to the
argument of \citet{gilmore2002}. It is notable that \citet{gilmore2002}
concluded, based on an analysis of their spectroscopic survey of some
2000 F/G stars located 0.5-5 kpc above the Galactic plane, that the
stars they proposed to originate in a shredded satellite exhibited a
large rotational velocity lag with respect to the thin/thick disk, on
the order of 100 km s$^{-1}$. Previous analyses for potential MWTD stars
generally did not consider stars with such a large lag as likely
candidate disk-like stars. Large lags for accreted MWTD stars may in
fact be expected, as argued by \citet{villalobos2009}. Clearly, care must
be exercised in the selection of potential MWTD stars before attempting
to discern their kinematic and chemical properties. 

\subsection{The Nature and Role of the MWTD in the Context of the Milky
Way}

The preponderance of evidence acquired prior to 2009 suggested that a
MWTD component exists, although doubts remained as to its level of contribution
to the numbers of metal-poor stars in the Solar Neighborhood, as well
as regarding its detailed kinematical behavior and relationship to the
canonical thick-disk component, and to the halo. In the period since
2009, a substantial volume of work has been carried out, making use of
large samples of stars with medium-resolution ($R \sim 2000$)
spectroscopy obtained from a variety of surveys, in particular the
Sloan Digital Sky Survey \citep[SDSS;][]{york2000}, and its Galactic
extensions, the Sloan Extension for Galactic Exploration and
Understanding \citep[SEGUE-1; ][]{yanny2009} and SEGUE-2 (C. Rockosi et
al., in preparation), as well as higher-resolution ($R \sim 7500$) data
from the Radial Velocity Experiment\citep[RAVE;][]{steinmetz2006}, and
other sources. A partial list of these works includes: 
\citet{carollo2010}, \citet{ruchti2010,ruchti2011},
\citet{kordopatis2011,kordopatis2013a,kordopatis2013b},
\citet{lee2011}, \citet{bovy2012a,bovy2012b,bovy2012c}, \citet{carrell12},
\citet{cheng2012a,cheng2012b},
\citet{pasetto12}, \citet{adibekyan2013}, \citet{boeche13a,boeche13b}, 
\citet{haywood13}, \citet{jayaraman2013}, \citet{bensby14}, and \citet{minchev14}.

During this period, our appreciation of the complexity of the halo has
also increased. \citet{carollo2007}, \citet{carollo2010}, and
\citet{beers2012} have presented the case that this system is
well-described in terms of an inner-halo and outer-halo population --
terms that we shall use in what follows\footnote{We note for
completeness that an alternative view has been expressed by
\citet{schoenrich2011}.}.  Additional evidence supporting the existence
of (at least) a dual halo has come from recognition that the frequency
of carbon-enhanced metal-poor (CEMP) stars that can be kinematically
associated with the outer-halo component is roughly twice that of the
inner-halo component \citep{carollo12}, analysis of the metallicity
distribution function (MDF), in combination with the motions, of local
halo stars by \citet{an13}, the apparent preference for stars of the
CEMP-$s$ sub-class (those exhibiting $s$-process-element
over-abundances) to be associated with the inner-halo component, while
stars of the CEMP-no sub-class (those exhibiting no neutron-capture
over-abundances) are more likely associated with the outer-halo
component \citep{carollo14}, and analysis of the in-situ change of the
halo system MDF with distance for a large sample of F-turnoff stars from
SDSS \citet{allende14}.

Furthermore, \citet{morrison2009} have used a sample of some 250 stars
with very well-determined kinematical properties to argue for the
presence of a new component of the local halo, with an axial ratio $c/a
\sim 0.2$ (similar in flattening to the thick disk) and populated by
stars with $-1.5 < $ [Fe/H] $ < -1.0$, which is, however, {\it not}
rotationally supported. The potential confusion of such stars with MWTD
candidates is obvious, due to their proximity to the Galactic plane.

These works have raised new and interesting questions concerning the
nature of the formation and evolution of both the disk and halo systems.
In the discussion of the MWTD, there seems to be no concensus yet as to
what the inter-relationships are between it, the canonical thick disk,
and the thin disk. Are they independent and discrete sub-systems? What
have been the roles (if any) of major mergers, the formation of the thin
disk from an early thick disk, the heating of a pre-existing thin
stellar disk by minor mergers, and/or efficient radial migration of
stars in the plane of the disk? We call the reader's attention to the
inciteful discussion by \citet{haywood13} of the apparently disparate
results concerning the nature of the thick/thin disks arising from
several recent studies. The issues being considered are clearly complex,
even in the face of high-quality data, and subtleties of the approaches
used and conclusions reached ensure that we have not yet arrived at a
widely accepted view.

\subsection{Scope of Present Investigation}

The focus of this paper is considerably narrower. Here, we
re-investigate the original sample of \citet{bidelman1973} discussed by
NBP, in order to resolve whether or not it includes substantial numbers
of stars that could be considered members of the MWTD population. New
high signal-to-noise medium-resolution ($R \sim 3000$) spectroscopy has
been obtained for some 300 stars of the NBP sample. Roughly one-third of
this sample now has available high-resolution spectroscopic
determinations of [Fe/H] (and other physical parameters) from the
literature, which we employ to carry out a calibration of the physical
parameter estimates obtained by a ``non-SEGUE'' version of the SEGUE
Stellar Parameter Pipeline \citep[SSPP; originally described
by][]{lee2008a}, and referred to as the n-SSPP. We also report estimates
of ``carbonicity,'' [C/Fe], calibrated with respect to some 50 of the NBP
stars with previous high-resolution spectroscopic determinations of this
ratio reported in the literature, supplemented by a number of additional
stars with available high-resolution determinations. The results of this
calibration effort will be used in a number of future investigations
based on non-SEGUE spectroscopic data. We then combine available radial
velocities, accurate proper motions from the Hipparcos and Tycho-II
catalogs, and the newly refined spectroscopic estimates of [Fe/H] to
consider the presence of a MWTD in this sample. The resulting
determinations of kinematic estimates represent a substantial
improvement in the space velocities derived by NBP, sharpening the
picture of the stellar populations obtained from these data, and
confirming the suggestion of NBP that the ELS paradigm is an
oversimplification of the manner in which the Milky Way formed.

This paper is outlined as follows. Our new observations are described in
Section~\ref{obs}, where we also discuss the determination of radial
velocities and spectroscopic line-strength indices for the program
stars. Section~\ref{atmpar} describes the techniques used to obtain estimates
of the stellar atmospheric parameters (\teff, \logg, \feh), as well as
[C/Fe], for these stars. A comparison of the newly derived metallicity
estimates with available high-resolution abundance results drawn from
the recent literature and with the original DDO photometry-based
estimates of NBP is presented in this section as well. We then make use of
these comparisons to carry out a calibration of the n-SSPP. Distance
estimates and proper motions for our sample stars are described in
Section~\ref{dis}. In Section~\ref{kin}, we use these data to perform a new
kinematic analysis of the NBP sample, and compare with the kinematics
derived from a similar sample of RAVE stars described by
\cite{ruchti2011}. A brief discussion of the implications of our new
results is presented in Section~ \ref{final}.  

\section{Spectroscopic Observations, Available Photometric Measurements,
and Derivation of Radial Velocities and Line Indices}
\label{obs}

\subsection{Details of Spectroscopic Observations}

During several observing runs conducted in January 1996, December 1996,
and June 1997, optical spectra for a total of 302 stars from the
subsample of 309 Bidelman \& Maconnell weak-metal candidates studied by
NBP (hereafter referred to as the B\&M sample) were obtained with the
Siding Spring Observatory 2.3m telescope, using the Double Beam
Spectrograph. These spectra covered the wavelength interval 3800 {\AA}
$\leq \lambda \leq$ 4500 {\AA}, with a resolving power of $R \sim 3000$,
similar to that obtained during the course of previous work on follow-up
spectroscopy by \cite{Norris1999} of metal-poor candidates selected from
the HK survey of \citet{beers1985,beers1992}. However, because the stars
in the current program are, in general, quite bright ($6.7 < V < 11.0$),
high-quality spectra (with S/N $> 100$ per resolution element) could be
obtained in reasonably short integration times. A total of 383 spectra
were obtained, including a number of stars with repeated measurements.
These spectra were reduced using standard procedures for flat-fielding,
extraction, and wavelength calibration (based on arc-lamp exposures
taken immediately before or after each science spectrum), using the
FIGARO software \citep{shortridge1993} and ancillary FORTRAN routines.
No attempt to spectro-photometrically calibrate the spectra was made.

\begin{figure}[!ht]
\epsscale{1.15}
\plotone{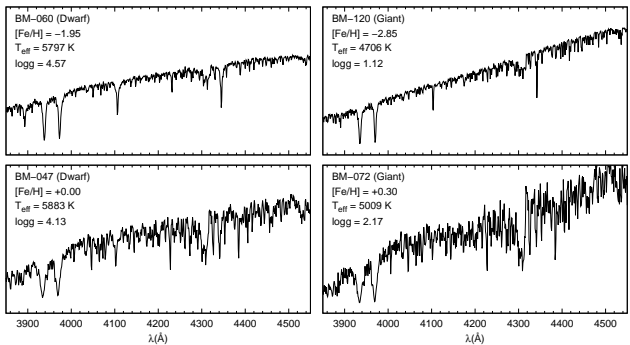}
\caption{Example medium-resolution ($R = 3000$) spectra for four of our program
stars obtained with the SSO 2.3m telescope. The left-hand column of
panels are main-sequence dwarfs, while the right-hand columns are giants
(and are classified as such by both the taxonomy of NBP and the present
analysis). The stellar atmospheric parameters from our analysis are
indicated in the legend of each spectrum.}
\label{spectra}
\end{figure}

Figure \ref{spectra} provides examples of the medium-resolution
spectra obtained. Also shown on the plots are the stellar atmospheric
parameters, determined using the methods described below
(Section~\ref{atmpar}). The left column of panels shows two examples
of stars classified as dwarfs by NBP (using taxonomy based on DDO
photometry), as well as by our own spectroscopic analysis. The
upper-left panel is the spectrum of a metal-poor dwarf (BM-060 =
CD-48:1741), while the lower-left panel is the spectrum of a dwarf
with solar metallicity (BM-047 = BD-13:959). The right column of
panels provides examples of spectra for stars classified as giants (by
both NBP and the present work). The upper-right panel is the spectrum
of a metal-poor giant (BM-120 = HD 84903), while the lower-right
panel is the spectrum for a giant with metallicity slightly above
solar (BM-072 = HD~40361).

\subsection{Broadband Photometry and Reddening Estimation}

The NBP study provided broadband $V$ magnitudes and $B-V$ colors for the
majority of our program objects. We have checked the SIMBAD database for
consistency with other measurements, and in a few cases replaced the
values listed by NBP with what we judged to be an improved set of
information. In some cases, photometry was not available from NBP. For
these stars, we adopted values provided in the SIMBAD database. The
results are listed in Table~\ref{tab1}. In this table, column (1) lists
the star names used by NBP, in the format BM-XXX, while column (2) lists
a more commonly used name for the star (e.g., BD, CD, HD, etc.). Columns
(3) and (4) are the Galactic longitude and latitude in decimal degrees,
respectively. Columns (5) and (6) list the adopted $V$ magnitude and
$B-V$ colors, respectively. Near-IR $JHK$ photometry is available for
the bulk of our sample, based on results from the 2MASS catalog
\citep{skrutskie2006}. The $J$ magnitude and $J-K$ colors reported by
2MASS for stars without flags indicating potential problems in the
listed values are given in columns (7) and (8), respectively. 

In order to obtain absorption- and reddening-corrected estimates of
the magnitudes and colors, respectively, we initially adopt the
\citet{schlegel1998} estimates of reddening, listed in column (9) of
Table 1. We have applied corrections to these estimates for objects
with reddening greater than $E(B-V)_S$ = 0.10, as described by
\citet{beers2002}. The corrected reddening estimates, $E(B-V)_A$, are
listed in column (10) of Table \ref{tab1}. The final reddening
estimates must be obtained in conjunction with the distance estimates,
obtained as described below, in order to properly account for the
amount of foreground reddening suffered by each star. The final
estimated reddenings, $E(B-V)_F$, are listed in column (11). Note that
about 20\% of our sample stars are located at low Galactic latitudes,
$|b| < $ 10$^{\circ}$, for which the reddening estimate along the line
of sight to a star is unreliable. For these stars we simply set the
reddening estimate to zero for the initial parameter
analysis\footnote{The n-SSPP employs multiple approaches, some of which
use spectroscopic-only input information, which provides robustness in
the parameter estimates in spite of spurious reddening estimates.}.

\subsection{Measurement of Radial Velocities and Line Indices}
\label{RV}

Our program stars, and the number of medium-resolution spectra
obtained for each star, are listed in the first two columns of Table
\ref{tab2}.  Radial velocities were measured for our program objects
using the line-by-line and cross-correlation techniques described in
detail by \citet{beers1999}, and references therein. The spectral
resolution is similar to that obtained for the majority of the HK
survey follow-up, so we expect that the measured radial
velocities should be precise to the same level (or better, given the
higher signal-to-noise of our present spectra), on the order of 7-10
km s$^{-1}$ (one sigma).  Heliocentric radial velocities
obtained from the medium-resolution spectra for our program stars,
RV$_M$, are listed in column (3) of Table \ref{tab2}.

\begin{figure}[!ht]
\epsscale{1.15}
\plotone{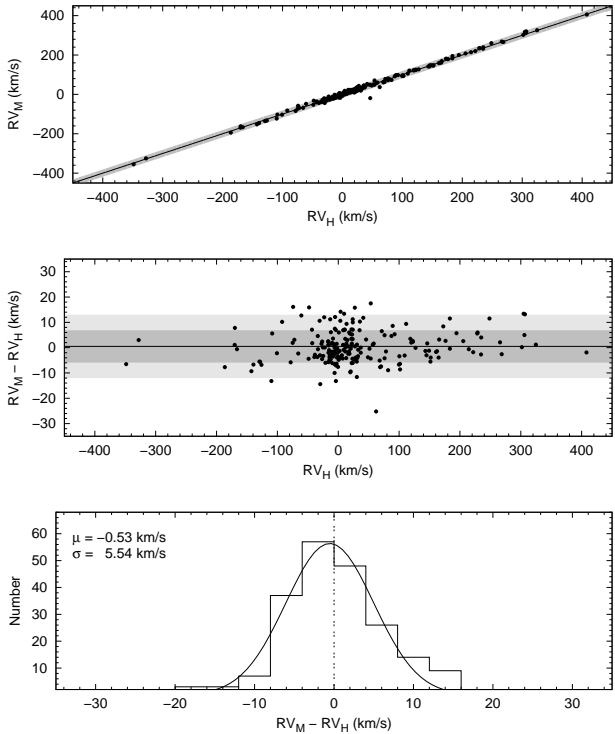}
\caption{Upper panel: Comparison between the radial velocities of our program
stars, determined from the medium-resolution spectra using the
techniques described by \citet{beers1999}, and those reported in the
literature from analyses of high-resolution spectra. The solid line is
the one-to-one line, and the shaded area represents a 3-$\sigma$ interval
around this line (where $\sigma$ represents the scatter in the residuals
shown in the lower panel, 5.5 \kms. Middle panel: Residuals between the
medium-resolution and high-resolution radial velocities, as a function
of the high-resolution values. The horizontal solid line is the average
of the residuals, while the darker and lighter shaded areas represent
the 1-$\sigma$ and 2-$\sigma$ regions, respectively. Lower panel:
Histogram of the residuals in the radial-velocity determinations. The
values of the mean offset and scatter are the parameters from the
Gaussian fit shown.}
\label{rv}
\end{figure}

Roughly one-third of our program objects have had radial
velocities determined from high-resolution spectroscopic studies
(available results are provided in column (4) of
Table~\ref{tab2}, RV$_{H}$). The upper two panels of Figure \ref{rv}
compare RV$_M$ with those obtained from the independent high-resolution
observations. As can be appreciated from inspection of this figure,
there is generally excellent agreement. A maximum likelihood fit to the
residuals in radial velocity, as shown in the lower panel of Figure
\ref{rv}, indicates that the rms scatter is only on the order of 5.5 km
s$^{-1}$. Assuming that the (combined) high-resolution radial velocities
from the literature have a precision on the order of 2.0 km s$^{-1}$,
the external errors in our medium-resolution radial-velocity
determinations appear to be no worse than about 5 km s$^{-1}$, 
slightly better than expected.  This represents a factor of two
improvement in the precision of the radial velocities reported by NBP.

Although we do not employ them for obtaining stellar metallicity
estimates in the present paper, we have measured a set of spectral
indices that have been used for this and other purposes in previous
papers (e.g., alternative schemes for estimation of de-reddened colors
based on Balmer-line strengths, [C/Fe] estimates based on the CH
$G-$band line index, etc.). Since these may prove useful in the future,
and are very well-measured in our high-S/N medium-resolution spectra, we
describe their determination below.

For each star, the measured (geocentric) radial velocities were used to
place a set of fixed bands for the derivation of line-strength indices,
which are pseudo-equivalent widths of prominent spectral features. We
employ a subset of the bands listed in Table \ref{tab1} of
\citet{beers1999}{\footnote{The indices KP, HP2, GP, and HG2 measure the
strength of the Ca II K line, hydrogen H$\delta$, the CH $G-$band, and
hydrogen H$\gamma$, respectively. A complete discussion of the choice of
bands and the ``band-switching'' scheme used to determine the indices
are provided in this reference as well.}.

Line indices for prominent spectral features for each of our program
stars are reported in columns (5)-(8) of Table~\ref{tab2}. A number of
our program stars had more than one spectrum obtained during the
course of the follow-up observations. From a comparison of the stars
with repeated measurements, we estimate that errors in the line
indices on the order of 0.1\,{\AA} are achieved. In order for a
line-index measurement to be considered a detection, we require that
the derived indices be above a minimum value of 0.25\,{\AA}. Indices
that failed to reach this minimum value are indicated in the 
Table~\ref{tab2} as missing data.

In addition to the line-strength indices, we have measured an
autocorrelation function index for each star, as described in detail in
\citet{beers1999}, and references therein.  We actually make use of the base-10 
logarithm of this index, hence it is referred to as LACF (listed in
column (9) of Table~\ref{tab2}). The LACF index quantifies the strength
of the multitude of weak metallic lines that are present in each
spectrum, and provides an additional indicator of the overall abundance.
This index is of particular use for cooler and/or metal-rich stars,
where the KP-index technique for inference of stellar metallicity
approaches saturation. 

\section{Stellar Atmospheric Parameters and Carbon Abundance Ratios}
\label{atmpar}

Stellar atmospheric parameters for our program stars were determined
using the n-SSPP, a modified version of the SEGUE Stellar Parameter
Pipeline \citep[SSPP; see][for a detailed description of the procedures
used]{lee2008a,lee2008b, allende2008, lee2011, smolin2011}. The n-SSPP
is a collection of routines for the analysis of non-SDSS/SEGUE data that
employs both spectroscopic and photometric ($V_0$, $(B-V)_0$, $J_0$ and
$(J-K)_0$) information as inputs, in order to make a series of estimates
for each stellar parameter\footnote{If both sets of $V,B-V$ and $J, J-K$
photometry are available, they are used, but the n-SSPP can operate well
with one or the other. Even if no photometric measurements are
available, the n-SSPP can, in most cases, produce viable stellar
parameter estimates (but not distance estimates, which require an input
apparent magnitude).}. Then, using $\chi^2$ minimization matching
techniques within dense grids of synthetic spectra, and averaging with
other techniques as available (depending on the wavelength range of the
input spectra; see Table 5 of Lee et al. 2008a), the best set of values
is adopted. For the SSPP, internal errors for the stellar parameters
are: 125~K for \teff, 0.25~dex for \logg, and 0.20~dex for \feh;
external errors are of a similar size. We might expect the external
errors in n-SSPP determinations to be somewhat larger, owing to the
generally more limited wavelength coverage and (in the present
application) lack of available $ugriz$ photometry. An empirical
determination of these errors for the n-SSPP is obtained below.

The spectra for our program stars do not reach as far redward as
SDSS/SEGUE spectra (hence we cannot use as many of the independent
methods as the SSPP provides), and they are of slightly higher resolving
power. Thus, during the execution of the n-SSPP, our spectra were first
rebinned in order to match the resolving power of SDSS/SEGUE spectra
(i.e., to 1 {\AA} linear pixels). The n-SSPP stellar atrmospheric
parameter estimates are listed in columns (10)-(12) of Table \ref{tab2}
as Teff$_S$, logg$_S$, and [Fe/H]$_S$, respectively.

The n-SSPP has been modified recently in order to estimate
carbon-to-iron abundance ratios (carbonicity, \cfe), based on
spectral matching against a dense grid of synthetic spectra.
\citet{lee2013} describes in detail the procedures adopted to estimate
\cfe{} for SDSS/SEGUE spectra; these techniques, with different input
photometric information, also apply to the n-SSPP. Note that we have
recently expanded the carbon grid to reach as low as [C/Fe] $= -1.5$,
rather than the limit of [C/Fe] $= -0.5$ employed by \cite{lee2013}. As
shown by \citet{lee2013}, the precision of the carbonicity estimates are
better than $0.35$~dex for the parameter space and (generally lower) S/N
ratios explored by SDSS/SEGUE spectra. We expect similar (or improved)
results for application of the n-SSPP to our program spectra, which is
checked empirically below. Table~\ref{tab3} lists the medium-resolution
estimates of carbonicity, [C/Fe]$_S$, in column (2). Column (3)
indicates whether the listed measurement is considered a detection,
DETECT = ``D'', lower limit ``L'', or upper limit ``U'', and Column (4)
provides the correlation coefficient, CC, obtained between the observed
spectrum and the best-matching [C/Fe] from the model grids. For an
acceptable measurement of carbonicity, we demand DETECT = ``D'' and CC
$\ge 0.7$. See \citet{lee2013} for further discussion of these
quantities. There are two stars listed in Table~\ref{tab3} (BM-083 and
BM-284) which have CC less than this value; they are marked with a ``:''
in the DETECT column.
 
\subsection{Comparison to High-Resolution Spectroscopic Analyses}

External measurements of atmospheric parameters and carbon abundances
were obtained from various sources in the literature, including the
compilations of \citet{strobel2001}, \citet{saga2008}, and
\citet{frebel10}, as well as the
references listed in SIMBAD\footnote{http://simbad.u-strasbg.fr/simbad/}
and in the PASTEL catalogue 
\citep{soubiran2010}\footnote{http://vizier.u-strasbg.fr/viz-bin/VizieR?-source=B/pastel}.
In total, we found 170 measurements of \teff, 111 of \logg, 114 of
\feh, and 50 for \cfe.  Note that we have
only made use of estimates based on studies published since 1990.  A
straight average of all available estimates for these parameters was
taken (excepting a few instances where it appeared that a given
high-resolution estimate was clearly highly discrepant); the results are listed 
in columns (13)-(15) of Table~\ref{tab2} as Teff$_H$, logg$_H$, and [Fe/H]$_H$, respectively.
Column (4) of Table~\ref{tab3} lists the high-resolution estimates of
carbonicity, [C/Fe]$_H$, for our program stars, where available. Since
the range in carbonicity for our program stars with
available high-resolution determinations is relatively limited, and does
not include many stars with [Fe/H] $ < -2.0$, we have supplemented the
comparison sample by obtaining n-SSPP estimates of [C/Fe]$_S$ from the
medium-resolution SDSS/SEGUE spectra for 39 stars with available
high-resolution determinations from \citet{aoki2013}.

\begin{figure}[!ht]
\epsscale{1.15}
\plotone{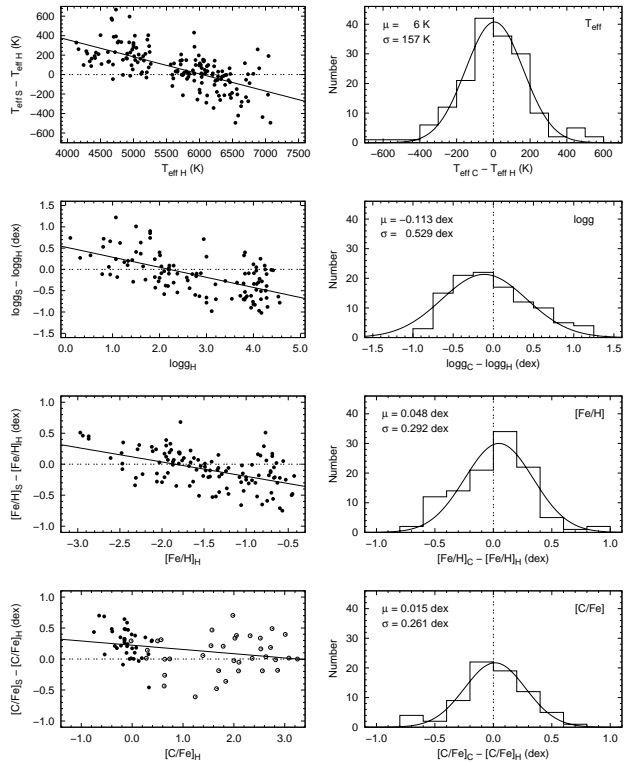}
\caption{Left panels: Differences between the atmospheric parameters 
and carbon-abundance ratios determined by the n-SSPP, Teff$_S$, logg$_S$,
[Fe/H]$_S$, and [C/Fe]$_S$, and the values
from analyses of high-resolution spectroscopy, Teff$_H$, logg$_H$,
[Fe/H]$_H$, and [C/Fe]$_H$ reported in the literature, as a function
of the high-resolution spectroscopic values.  Filled symbols refer
to our B\&M program stars, while in the bottom panel the open
symbols represent stars from Aoki et al.\ (2013), as described in the
text. The dashed lines show the linear functions (Equations
(1)-(4)) used to obtain corrections to the n-SSPP values, as described
in the text. Right panels: Histograms of the residuals between the
corrected n-SSPP and high-resolution parameters shown in the left
panels. Each panel also lists the average offset and scatter
determined from a Gaussian fit.}
\label{correct}
\end{figure}

Figure \ref{correct} illustrates, in the left-hand column of panels,
comparisons of the n-SSPP estimates Teff$_S$, logg$_S$, [Fe/H]$_S$,
and [C/Fe]$_S$ with the averaged high-resolution spectroscopic
results, Teff$_H$, logg$_H$, [Fe/H]$_H$, and
[C/Fe]$_H$.  The solid lines in these panels are linear fits to the
residuals in the difference between the medium- and high-resolution
results, as a function of the high-resolution determinations. We use these
fits to correct our n-SSPP estimates from the medium-resolution
spectra to come into better agreement with the external
high-resolution estimates, in the form:

\begin{eqnarray}
{\rm [Fe/H]}_C     = {\rm [Fe/H]}_S - ( -0.232 \cdot {\rm [Fe/H]}_S - 0.428 )  \\
{\rm Teff}_C   = {\rm Teff}_S -  ( -0.1758 \cdot {\rm Teff}_S + 1062)  \\
{\rm logg}_C       = {\rm logg}_S -  ( -0.237 \cdot {\rm logg}_S + 0.523 )  \\
{\rm [C/Fe]}_C    = {\rm [C/Fe]}_S - ( - 0.068 \cdot {\rm [C/Fe]}_S + 0.273) 
\label{eqcor}
\end{eqnarray}

The right-hand column of panels in Figure~\ref{correct} shows the
distribution of residuals between the corrected n-SSPP estimates
(Teff$_C$, logg$_C$, and [Fe/H]$_C$, listed in columns (16)-(18) of Table
\ref{tab2}, and [C/Fe]$_C$, listed in column (6) of Table~\ref{tab3}) and
their corresponding high-resolution values. Note that, with the
exception of a few individual stars, the agreement is quite
satisfactory. Maximum-likelihood Gaussian fits to the
distributions of residuals between these various estimates are shown in
the right-hand column of panels. Taking into account the expected errors
in the (non-uniformly analysed) high-resolution literature estimates of
the effective temperature and surface gravity (100~K and 0.35 dex,
respectively), we conclude that the external accuracies of Teff$_C$ and
logg$_C$ are on the order of 125~K and 0.4 dex, respectively. The
zero-point offsets of the n-SSPP estimates are acceptably small, on the
order of 6~K and 0.1 dex for \teff$_C$ and logg$_C$, respectively. The
rather large external error in the surface gravity estimate is perhaps
not surprising, since the spectra do not extend sufficiently redward to
include the particularly gravity sensitive \ion{Mg}{1} lines at $\sim
5180$\,{\AA}. Assuming that the expected errors in the literature
estimates of metallicity and carbonicity are on the order of 0.20 dex (which
may be generous), the external errors in [Fe/H]$_C$ and
[C/Fe]$_C$ are both $\sim 0.20$ dex.

\subsection{Comparison to the NBP DDO-based Estimates}

We now examine a comparison of our presently determined spectroscopic
metallicity estimates with the DDO photometry-based estimates of
metallicity given by NBP, and listed as [Fe/H]$_N$ in column (19) of
Table~\ref{tab2}. The upper panels of Figure~\ref{norris_comp2} show the
complete sample, while the middle and lower panels for stars classified
as dwarfs and giants by NBP, respectively.

The luminosity classes for our program stars are listed in column (20),
in the form TYPE$_{N/S}$, where ``N'' indicates those assigned by NBP,
while ``S'' indicates the classes assigned by the n-SSPP. The stars
classified as dwarf, subgiant, and red giant map directly onto the
classes considered by \citet{beers2000} (note that we do not discriminate
between subgiants and giants in the n-SSPP; they are all classified as
giants). We consider the stars NBP classified as blue dwarfs to be
main-sequence turnoff (TO) stars, while the blue giant and red
horizontal-branch (RHB) classes are considered as field
horizontal-branch (FHB) stars. The UV bright stars are considered to be
giants. Note that the classifications are commensurate, in most cases,
although they differ for 43 stars (labeled with a ":" in column (20)).
The great majority of these conflicting classifications (33 of 43)
occur for stars with \teff$_C$ $\ge 6000$~K, the region where it becomes
difficult to distinguish dwarfs from giants close to the main-sequence
turnoff. We proceed with our analysis under the assumption the n-SSPP
classification is superior, and make use of it for cases where the
luminosity class is in doubt (except where noted).

\begin{figure}[!ht]
\epsscale{1.15}
\plotone{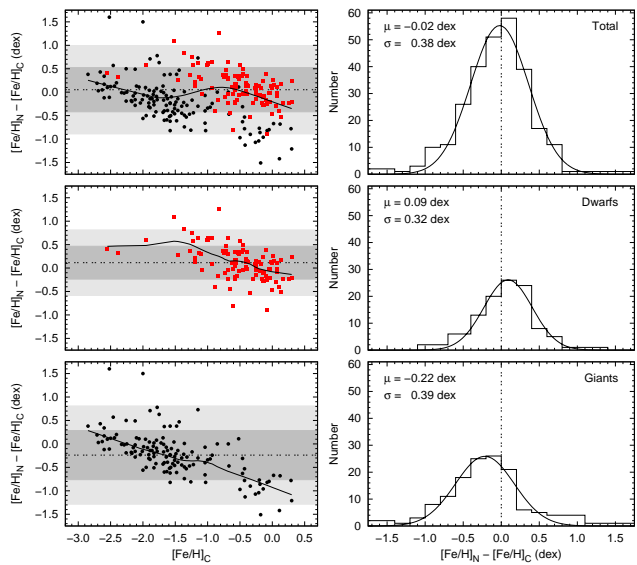}
\caption{Comparison between the DDO photometry-based metallicities from NBP with
our final metallicity estimates, [Fe/H]$_C$, for the entire sample
(upper panels), dwarfs (middle panels), and giants (lower panels). Note
that, for the purposes of this comparison, the taxonomy of NBP was used.  The
left-hand column of panels shows the residuals as a function of
[Fe/H]$_C$, with the average (dashed horizontal line), while the darker
and lighter shaded areas represent the 1-$\sigma$ and 2-$\sigma$
regions, respectively. The solid line is a {\it loess} (locally
weighted regression) line. The right panels show Gaussian fits to the
residuals. Note the clear offset to lower metallicities reported by NBP for the giants
with metallicities [Fe/H]$_C > -1.5$ (see text).}
\label{norris_comp2}
\end{figure}

The average offset and scatter in the metallicity residuals, shown in
Figure~\ref{norris_comp2} for the complete sample, are ($\mu$, $\sigma$)
= ($-0.02$ dex, 0.38 dex), while the values for the dwarfs and giants
(as classified by NBP) are ($+0.09$ dex, 0.32 dex) and ($-0.22$ dex,
0.39 dex), respectively. The solid lines in the left-hand panels
indicate locally weighted regression ({\it loess}) lines that trace the
data. As can be appreciated by inspection of the middle panels of
Figure~\ref{norris_comp2}, the dwarfs exhibit general agreement with our
spectroscopic metallicity estimates, with a tendency for [Fe/H]$_N$ to
be somewhat higher than [Fe/H]$_C$ for metallicities [Fe/H]$_C < -1.0$.
By way of contrast, the [Fe/H]$_N$ for giants in the lower panels are
clearly too low, compared to [Fe/H]$_C$, for [Fe/H]$_C > -1.5$, and
somewhat higher than [Fe/H]$_C$ for [Fe/H]$_C < -1.5$. It is the higher
metallicity stars that disagree in the same sense described by
\citet{anthony1994} and \citet{ryan1995}, and which were the source of 
concern for the validity of the original claim for a
MWTD. What remains to be shown is whether the existence of a MWTD is
supported, or refuted, by the analysis we carry out below, using our
improved estimates of metallicity and refined kinematics.

For the remainder of the paper, we drop the subscripts on [Fe/H]$_C$ and
[C/Fe]$_C$, and simply refer to our adopted metallicity
and carbonicity estimates as [Fe/H] and [C/Fe], respectively.

\section {Distance Estimates and Proper Motions}
\label{dis}

\subsection {Distance Estimates}
\label{dist}

Distances to individual stars in this sample are estimated using the
$M_V$ vs. $(B-V)_0$ relationships described by \citet{beers2000}. These
relationships require that the likely evolutionary state (luminosity
class) of a star be specified. For this, we make use of the taxonomy
assigned by the n-SSPP (which assigns types according to the observed
surface gravity estimate), the most recent discussion of which is
provided by \citet{beers2012}. Note that this includes the reassignment,
as necessary, of stars classified as main-sequence turnoff stars into
dwarfs or giants when they otherwise appear in physically impossible
positions in the color-magnitude diagram. See \citet{beers2012} for
additional details.

Once types are assigned, distance estimates can be obtained in a
straightforward manner. The estimates have to be iterated, because both
$V_0$ (and therefore the distance estimate) and $(B-V)_0$ depend on the
adopted reddening. Although the $M_V$ vs. $(B-V)_0$ relationships depend
on the metallicity as well, the change in metallicity with small
alterations in reddening has little effect. With only a few iterations
we obtain consistent estimates of the final reddening, $E(B-V)_F$
(listed in column (11) of Table \ref{tab1}), and the photometric distance
estimate, $D_{\rm pho}$, listed in the fourth column of
Table~\ref{tab4}. Based on previous tests of this approach (e.g.,
Beers et al. 2000, 2012), we expect the photometric distances to be
precise to on the order of 10-20\%. For most instances, we apply a
distance uncertainty of 15\%, although in a few cases, larger
uncertainties were adopted in order to reflect uncertainties in the
determination of reddening corrections.

All but four stars among our program objects have parallaxes available
from the Hipparcos astrometric catalog \citep{esa1997,leeuwen2007}.
These parallaxes, and their associated errors, are listed in columns (2)
and (3) of Table~\ref{tab4}. 

To assess the reliability of our photometric distance estimates,
Figure~\ref{dist1} shows the comparison between distances based on
photometry, $D_{\rm pho}$, with the distances based on Hipparcos
parallaxes, $D_{\rm HIP} = 1/\pi_{\rm HIP}$, for the four different
luminosity classes of the targets. Note that this plot only includes
stars for which accurate trigonometric and photometric distance
estimates are available (that is, we exclude stars with trigonometric
parallaxes having $\sigma_{\pi_{\rm HIP}}/\pi_{\rm HIP} > 0.20$, listed
in column (6) of Table~\ref{tab4}, or located at the lowest latitudes,
$|b| \le 10^{\circ}$, where reddening, and hence extinction, to a given
star is highly uncertain. As can be appreciated by inspection of
Figure~\ref{dist1}, the relationship between our derived photometric and
Hipparcos distances is close to the one-to-one line. Only the few stars
with distances greater than about 200 parsecs exhibit significant
scatter. 

\begin{figure}[!ht]
\epsscale{1.15}
\plotone{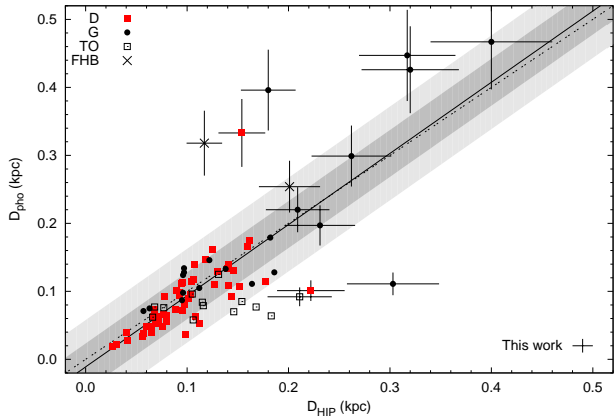}
\caption{Comparison of the photometrically estimated distances,
$D_{\rm pho}$, with the trigonometric distance estimates, $D_{\rm
 HIP}$, for stars with sufficiently accurate Hipparcos parallaxes
($\sigma_{\pi_{\rm HIP}}/\pi_{\rm HIP} \le 0.20$). The dashed line
is the one-to-one line, while the solid line is a robust regression
fit to the data. The darker and lighter shaded areas
represent the 1-$\sigma$ and 2-$\sigma$ regions about the linear
fit, respectively, based on a Gaussian fit to the residuals. The
error bar in the lower right corner of the plot is the typical error
for stars with distances less than 200 pc. For stars with distances
(either photometric or astrometric) greater than 200 pc, individual
error bars are shown. }
\label{dist1}
\end{figure}

For the purpose of our kinematic analysis below, we adopt distance
estimates based on trigonometric parallaxes, where we judge them to be
sufficiently accurate (parallaxes satisfying $\sigma_{\pi_{\rm
HIP}}/\pi_{\rm HIP} \le 0.20$, and greater than zero). Otherwise, we
adopt the derived photometric distance estimate. The final
adopted distances, $D_{\rm ado}$, are listed in column (7) of
Table~\ref{tab4}.

\subsection{Proper Motions}
\label{motions}

Proper motions for all of our program stars are available from the
Hipparcos \citep{esa1997,leeuwen2007}, Tycho-1
\citep{esa1997,hog1998}, or Tycho-2 \citep{esa1997,hog2000} catalogs,
with average precisions of 1.25 mas yr$^{-1}$ in $\mu_{\alpha}$ (taking
the $\cos \delta$ term into account) and 1.03 mas yr$^{-1}$ in
$\mu_{\delta}$. The precision of the presently available proper motions
represent substantial improvements over those used by NBP. Furthermore,
while all of our program stars have proper motion estimates, only about
one-third of the stars in the NBP catalog had this information
available. Columns (8) and (9) of Table~\ref{tab4} list the proper
motions for our program stars, while columns (10) and (11) present their
associated errors. The final column lists the identifier of the star in
the Hipparcos or Tycho catalogs.
 
\section{Kinematic Analysis of the B\&M Sample}
\label{kin}

In this section we examine the kinematic properties of the
B\&M sample of stars studied by NBP.

Figure~\ref{samp} shows the distribution of the absorption-corrected
$V_0$ magnitudes, de-reddened $(B-V)_0$ colors, adopted distances,
$D_{\rm ado}$, and estimates of metallicities, [Fe/H], for our sample of program stars.
As is immediately clear from inspection of this figure, this is a very local
sample of stars, with $\sim 90$\% of the stars located within 1 kpc from
the Sun. Nevertheless, because of the metallicity bias in the original
selection, some 70\% have [Fe/H] $\le -0.5$, suitable for exploration of
the thick disk, MWTD, and inner-halo population. There are not large
numbers of stars in this sample with [Fe/H] $< -2.0$ ($\sim$ 10\% of the
sample), which limits its utility for examination of the outer-halo
population.

\begin{figure}[!ht]
\epsscale{1.15}
\plotone{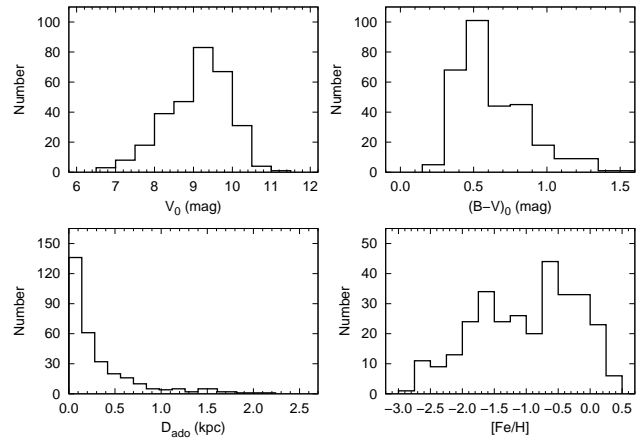}
\caption{Distributions of (a) absorption-corrected V$_0$ magnitudes, (b) de-reddened $(B-V)_0$ colors, (c)
adopted distance estimates, $D_{\rm ado}$, and (d) metallicity estimates,
[Fe/H], for our program stars.}
\label{samp}
\end{figure}

\subsection {Determination of Space Motions and Orbital Parameters}

It is our intention to obtain the most precise kinematics for our
program stars, taking advantage of the refinements that more modern data
have provided in the 28 years since the publication of the first paper
in this series. 

Adopted distances are determined as described in subsection~\ref{dist}
above. Proper motions, almost all taken from the Hipparcos catalog, are
described in subsection~\ref{motions} above. From Table~\ref{tab2}, we
adopt the high-resolution spectroscopic determinations of radial
velocities, RV$_H$, where available, which are expected to have
precisions of 2 \kms\ or better. When not available, we make use of
the determinations based on medium-resolution spectroscopy, which were
demonstrated in subsection~\ref{RV} to exhibit precisions on the order
of 5 \kms.     

We now derive the space motions and orbital parameters of our program
stars, following the procedures described by \citet{carollo2010}.

Corrections for the motion of the Sun with respect to the Local Standard
of Rest (LSR) are applied during the course of the calculation of the
full space motions; here we adopt the values $(U,V,W)$=(9, 12, 7) \kms\
\citep{mihalas1981}. Note that we follow the convention that $U$ is
positive in the direction away from the Galactic center, $V$ is positive
in the direction of Galactic rotation, and $W$ is positive toward the
north Galactic pole. For the purpose of this analysis it is also
convenient to obtain the rotational component of a star's motion about
the Galactic center in a cylindrical frame, denoted as \vphi, and
calculated assuming that the LSR is on a circular orbit with a value of
220 \kms\ \citep{kerr1986}. Our assumed values of \rsun\ (8.5 kpc) and
the circular velocity of the LSR are both consistent with two recent
independent determinations of these quantities by
\citet{ghez2008} and \citet{koposov2009}. \citet{bovy2012} recently obtained
an estimate of the Milky Way's circular velocity at the position of the
Sun of V$_c$(\rsun) = $218 \pm 6$ \kms, based on an analysis of
high-resolution spectroscopic determinations from the Apache Point
Observatory Galactic Evolution Experiment \citep[APOGEE;
][]{majewski2010}, part of the Sloan Digital Sky Survey III
\citep[SDSS-III;][]{eisenstein2011}, which is also consistent with our
adopted value.

The orbital parameters of the stars, such as the perigalactic distance
(the closest approach of an orbit to the Galactic center), \rperi, the
apogalactic distance (the farthest extent of an orbit from the Galactic
center), \rapo, of each stellar orbit, and the orbital eccentricity,
$e$, defined as $e$ = (\rapo~-~\rperi)/(\rapo~+~\rperi), as well as
\zmax\ (the maximum distance that a stellar orbit achieves above or
below the Galactic plane), are derived by adopting an analytic
St\"ackel-type gravitational potential \citep[which consists of a
flattened, oblate disk, and a nearly spherical massive dark-matter halo;
see the description given by][ Appendix A]{chiba2000}, and integrating
their orbital paths based on the starting point obtained from the
observations.  

Table~\ref{tab5} provides a summary of the above calculations. Column
(1) provides the star names. Columns (2) and (3) list the positions of
the stars in the meridional ($R,Z$)-plane. The derived $UVW$ velocity
components are provided in columns (4)-(6); their associated errors
are listed in columns (7)-(9). Column (10) lists the
velocity projected onto the Galactic plane ($V_R$, positive in the
direction away from the Galactic center), while column (11) lists the
derived rotation velocity, \vphi. The derived \rperi\  and \rapo\
are given in columns (12) and (13), respectively.
Columns (14) and (15) list the derived \zmax\ and orbital eccentricity, $e$,
respectively.

\begin{figure}[!ht]
\epsscale{1.15}
\plotone{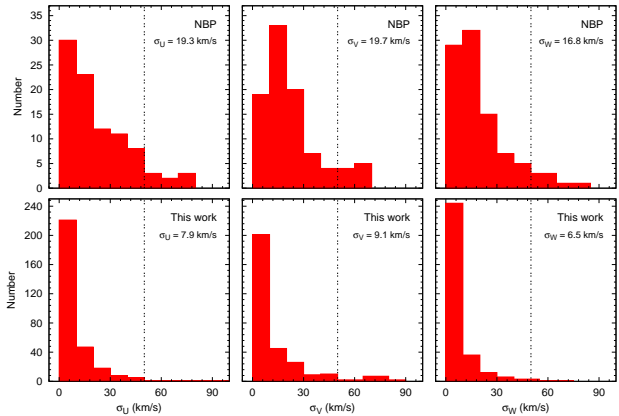}
\caption{Errors in the estimation of the local velocity components of
the space motions for stars in the B\&M sample, for the
stars reported by NBP (upper panels) and in the present work (lower
panels).  The vertical dashed lines at 50 \kms\ indicate the maximum
individual errors allowed for a given star to be included in the
subsequent kinematic analysis.  The legends provide the mean errors for
the accepted stars.  Note the marked decrease in the
errors for these quantities obtained by the present work.}
\label{euvw}
\end{figure}

Errors on our derived estimates of the individual components of the
space motions take into account an estimated 15\% error in the
photometric distances (individual errors in Hipparcos distances, when
adopted, are used), as well as the individual errors in the proper
motions and the adopted radial velocities (2 \kms\ for the
high-resolution determinations, 5 \kms\ for the medium-resolution
determinations). As expected, when compared to the previous results of
NBP, the derived errors in these quantities are much improved.
Figure~\ref{euvw} shows the distributions of these errors for both sets
of analyses. After removing the stars with individual estimated errors
in any one of the three components of space motion larger than 50 \kms\
from each of these samples (or which were dropped for other reasons),
the average errors for the B\&M sample are $\sigma (U, V, W)$ = (7.9,
9.1, 6.5) \kms. For NBP, the average errors were 2 to 2.5 times as high,
$\sigma (U, V, W)$ = (19.3,19.7,16.8) \kms. The large errors in the
individual space motions (and eccentricities) of some stars forced NBP
to rather severely trim their sample from which inferences could be made
about the nature of the underlying populations.

A total of 42 stars in our full sample of 302 stars are not used in the
kinematic analysis, because they are either missing one or more of the
input quantities used for the determination of their space motions, are
located at Galactic latitudes $|b| < $ 10$^{\circ}$ and had 
only photometric distances available (and hence uncertain estimates
of reddening), or had individual estimated errors in any one of the three
components of space motion larger than 50 \kms. Such stars are noted in
the final column of Table~\ref{tab5}, where the first digit of the INOUT
parameter set to ``0'' indicates that the star has been dropped from
subsequent kinematical analysis.

\subsection{Distributions of $UVW$, and \zmax\ vs. [Fe/H]}
 
Figure~\ref{uvw} presents the individual components of the space
motions, as a function of [Fe/H], for our program stars with accepted
kinematic estimates.  It is clear from inspection of this
diagram that there exists a ``core'' of stars with relatively high net
rotation and low velocity dispersion down to at least [Fe/H] = $-1.3$,
and possibly a little lower.  This immediately suggests the presence of
low-metallicity stars in the disk system, well below the mean abundance
typically associated with the canonical thick disk, on the order of
[Fe/H] = $-0.6$.  

\begin{figure}[!ht]
\epsscale{1.15}
\plotone{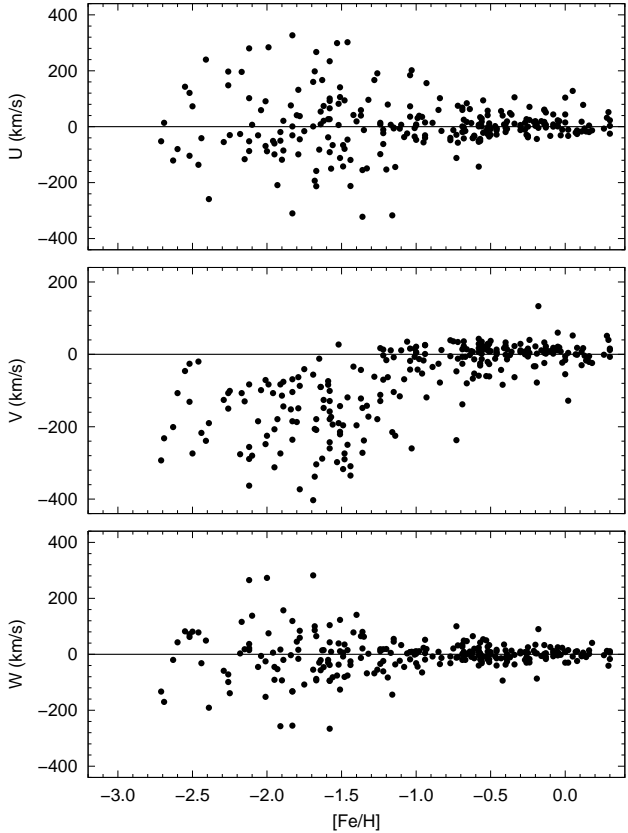}
\caption{Local velocity components for stars in the B\&M sample with
available $UVW$ estimates, as a function of metallicity, [Fe/H]. Note
the existence of stars with low velocity dispersions in their estimated
components down to at least [Fe/H = $-1.5$, and possibly a little lower.
Stars with errors in any of the individual derived components of motion
exceeding 50 \kms\ are excluded. }
\label{uvw}
\end{figure}

Since the great majority of our program stars are located within 1 kpc,
it is difficult to separate possible stellar populations on the basis of
vertical distance from the Galactic plane, As an alternative, we have
explored using the derived maximum distance from the plane, \zmax.
Figure~\ref{zmax_feh} shows the result of this exercise, where we have
plotted the base-10 log of \zmax\ as a function of [Fe/H]. A reference
line at \zmax\ = 3 kpc is shown. From inspection of this figure, it is
clear that significant numbers of stars with \zmax\ $> 3$ kpc are not
found for [Fe/H] $\ga -1.5$. Although some overlap between the
inner-halo population and the proposed MWTD population is unavoidable,
we expect the majority of the stars with \zmax\ $\le 3 $ kpc and in the
metallicity interval $-1.8 \le $ [Fe/H] $ \le -0.8$
\citep[following][]{carollo2010} to be associated with the MWTD.  
As is shown below, stars in this metallicity interval also exhibit a
lower net rotation than the stars of the canonical thick disk.

\begin{figure}[!ht]
\epsscale{1.35}
\plotone{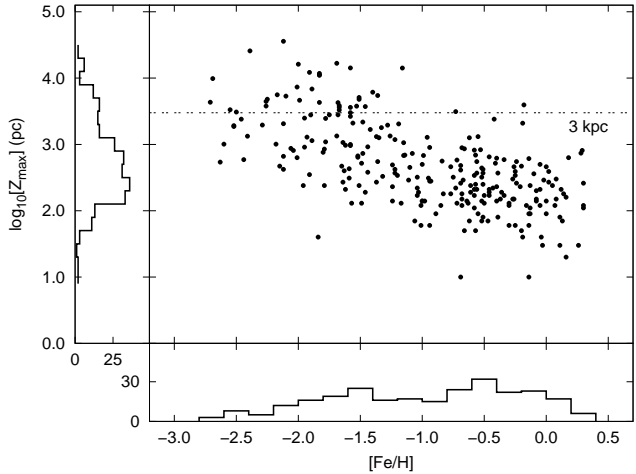}
\caption{Distribution of $Z_{\rm max}$, the largest distance above or below
the Galactic plane achieved by a star during the course of its orbit, as
a function of metallicity, for the stars in the B\&M sample. The
marginal distributions of each variable are shown as histograms. The
horizontal dashed line provides a reference at 3 kpc. Very few stars
with metallicity [Fe/H] $ > -1.5$ achieve orbits that reach higher than
this location. Note the logarithmic scale for $Z_{\rm max}$. Stars with
errors in any of the individual derived components of motion exceeding
50 \kms\ are excluded. }
\label{zmax_feh}
\end{figure}

\subsection{The [Fe/H] vs. Eccentricity Diagram}

One of the central arguments of NBP, that there exist numerous stars in
the B\&M sample (where none were found in the sample used by ELS to
support their classic monolithic collapse model) in the regime [Fe/H]
$\le -1.0$, $e \le 0.4$, can now be re-examined using our improved
kinematic results. The upper panel of Figure~\ref{feh_e} is a plot of
[Fe/H], as a function of orbital eccentricity, for the sample of B\&M
stars from NBP (their Figure 14), where we have used their derived
metallicities and eccentricities (restricted, as did they, to stars with
estimated errors in the eccentricity $\le 0.1$). The lower panel of the
figure shows the more comprehensive results for the B\&M stars from the
present analysis. All of our program stars with acceptable kinematic
determinations have estimated errors in eccentricity less than 0.1. In
both panels, for heuristic purposes, the red squares stand for red
giants (NBP, upper panel) and giants (this work, lower panels), while
black circles are used for all other objects. The two panels also show
the original box ($-2.0 \le$ [Fe/H] $\le -1.0$, $0.2 \le e \le 0.4$)
used by NBP to contrast the locations of the stars in the ELS sample
with their own. Note that there are roughly 10 stars in the NBP analysis
that fall in the box (plus or minus a few that are right at the edges of
the box) compared with none in the original ELS sample. As seen in the
lower panel, the present analysis now includes almost twice as many
stars in the same box.

\begin{figure}[!ht]
\epsscale{1.15}
\plotone{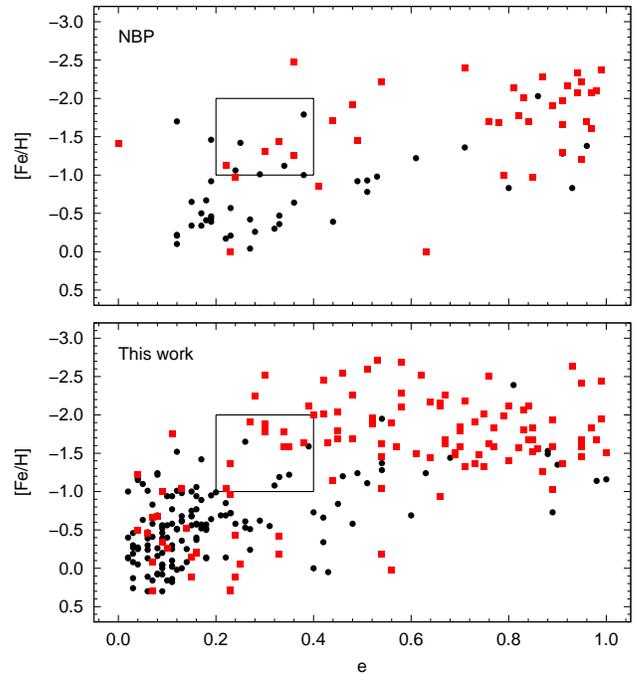}
\caption{Distribution of metallicity, [Fe/H], for stars in the B\&M
sample, as a function of derived orbital eccentricity,
$e$. In both panels the red squares stand for red giants
while black circles are used for all other objects. (Upper panel)
The sample studied by NBP (and adopting their metallicities and
eccentricities), which only included stars with eccentricities
having errors less than 0.1. The box indicates the region of
interest originally considered by NBP. (Lower panel) The present
sample, all of which have errors in eccentricity less than 0.1. In
addition to a much larger total sample, note that the box now
contains about twice as many stars as for the NBP sample. Stars with
errors in any of the individual derived components of motion
exceeding 50 \kms\ are excluded. }
\label{feh_e}
\end{figure}

Clearly, the addition of more stars with improved estimates of
metallicities and eccentricities in the present sample {\it strengthens}
the original thesis of NBP that the ELS model was incompatible with
results obtained for the B\&M sample of non-kinematically selected
stars. 

Of course, more modern analyses of even larger such samples
\citep[e.g.,][]{chiba2000,carollo2007,carollo2010} have come to similar
conclusions, and are consistent with the expectations from contemporary
hierarchical galaxy-assembly models. However, this idea had its
observational origin for field stars in the work of NBP\footnote {J.E.N.
acknowledges here the validity of the criticism of the NBP abundances
for red giants by \citet{anthony1994} and \citet{ryan1995}; for
historians of science, he notes that the discrepancy results from errors
in the limited DDO abundance calibration adopted for (absolutely)
fainter, more metal-rich giants. That said, inspection of the present
Figure~\ref{feh_e} shows that, with the present abundance calibration,
giants exist in the region where ELS (due to their selection criteria)
found no stars .}.

\subsection{The Toomre and Lindblad Diagrams}

The so-called Toomre diagram (a plot of ($U^2 + W^2)^{1/2}$, the
quadratic addition of the $U$ and $W$ velocity components, as a function
of the rotational component, $V$) and the Lindblad diagram (a plot of
the integrals of motion representing the total energy, $E$, and the
vertical angular momentum component, $L_Z$) are commonly used to
investigate the nature of the kinematics of stellar populations in the
Galaxy. Given the high quality of the estimated kinematics for the B\&M
sample, it is useful to examine these diagrams to glean any insight
we can from them.

For the sake of comparison, we have also derived these diagrams based on
the sample of RAVE stars reported by \citet{ruchti2011}\footnote{We have
updated the estimates of [Fe/H] and the derived $UVW$ used for the
Ruchti et al. sample based on revised information kindly supplied by G.
Ruchti (private communication), based on a newer version of the RAVE
pipeline outputs.}. The Ruchti et al. sample is of similar size, and
covers a similar metallicity range as the B\&M sample, but carries along
its own set of biases in the selection of the member stars \citep[see][
for a discussion]{ruchti2010}. In addition, the proper motions available
to Ruchti et al. are not as precise, in general, as those in our sample,
since few of their stars were included in the Hipparcos, Tycho, or
Tycho-2 catalogs.
  
Figure~\ref{toomre} shows the Toomre diagrams for the two samples. The
upper panel is the sample of Ruchti et al. (after removal of the 72 of
319 stars that are either missing the input quantities used for
determination of their space motions, or that have individual estimated
errors in any one of the three components of space motion larger than 50
\kms). Following the removal of these stars, the average errors in the
derived space motions are $\sigma (U, V, W)$ = (14.4, 15.6, 12.2) \kms,
about twice as high as from our analysis of the B\&M sample, but still
better than obtained from the NBP analysis.

\begin{figure}[!ht]
\epsscale{1.15}
\plotone{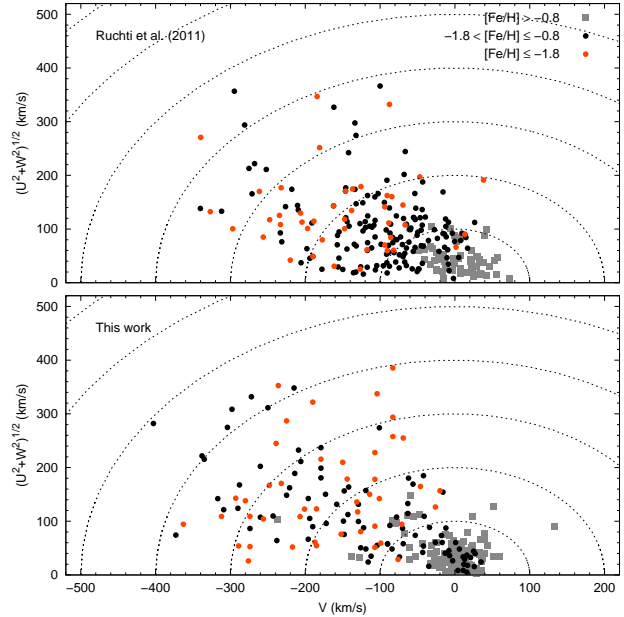}
\caption{Toomre diagram of ($U^2 + W^2)^{1/2}$ vs. V for
the stars with available $UVW$ velocity components, in three regimes of
metallicity, for stars the sample of RAVE stars from
\citet{ruchti2011} (upper panel) and in the B\&M sample (lower
panel). Note the presence of the intermediate-metallicity ($-1.8 <
[Fe/H] \le -0.8$) stars both inside and outside the region with low
($U^2 + W^2)^{1/2}$ and high orbital rotation
velocities (($U^2 + W^2)^{1/2} \lesssim 100$ \kms, $-100 < V < 100$
\kms). Stars with errors in any of the individual derived components
of motion exceeding 50 \kms\ are excluded. }
\label{toomre}
\end{figure}

The lower panel shows the Toomre diagram for the B\&M sample. The legend
indicates the metallicity intervals that are distinguished in the two
panels. These were chosen to roughly separate stars expected to belong
to the thick (or thin) disk ([Fe/H] $> -0.8$), the suggested MWTD ($-1.8
< $ [Fe/H] $\le -0.8$), and the halo system ([Fe/H] $\le -1.8$), taking
our guidance in selecting these intervals from \citet{carollo2010}. As
expected, the more metal-rich stars in both samples are primarily found
in the region with low ($U^2 + W^2)^{1/2}$} and high orbital rotation
velocities ($U^2 + W^2)^{1/2} \lesssim 100$ \kms, $-100 < V < 100$
\kms), while stars with intermediate metallicies are divided between
those inside and outside this region.

\begin{figure}[!ht]
\epsscale{1.20}
\plotone{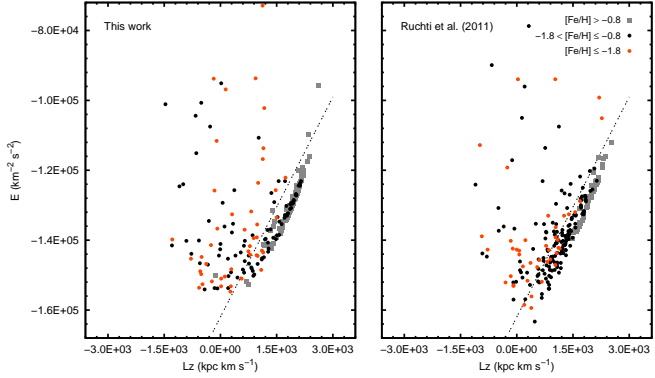}
\caption{Lindblad diagram of two integrals of motion, the total
energy, $E$, and the vertical angular momentum component, $L_Z$), in
three regimes of metallicity, for the B\&M sample (left panel) and for
the sample of RAVE stars from \citet{ruchti2011} (right panel). The
dashed line is drawn in order to separate likely members of
rotation-supported populations (see text). Note that the same line is
drawn in both figures. Stars with errors in any of the individual
derived components of motion exceeding 50 \kms\ are excluded.}
\label{lindblad}
\end{figure}

The Lindblad diagrams for these two samples of stars, calculated
following the prescription described by \citet{carollo14}, are shown in
Figure~\ref{lindblad}. As can be appreciated from inspection of the
left-hand panel, which applies to the B\&M sample, there is rather clear
separation of a region corresponding to a rotationally supported disk
system, indicated by the dashed line shown to guide the eye. The same
line is drawn in the right-hand panel, which applies to the Ruchti et
al. sample. The stars to the right of the line are expected to be
dominated by members of the disk system (thin, thick, and MWTD) while
those to the left of the line are likely to be dominated by inner-halo
members. Note that the separation of stars across this line is somewhat
less clear for the Ruchti et al. sample, presumably because of the
larger errors in the derived kinematics. Also note that this kinematic
division does not isolate only the more metal-rich stars, but in both
samples appears to include relatively large fractions of stars in the
intermediate metallicity interval, $-1.8 < $ [Fe/H] $\le -0.8$, and a
few stars with even lower metallicity.

\subsection{Distributions of \vphi~ for the B\&M and Ruchti et al. Samples}

\begin{figure}[!ht]
\epsscale{1.20}
\plotone{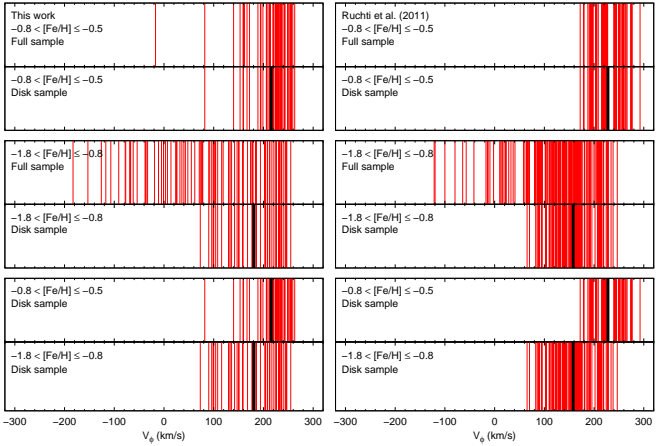}
\caption{Stripe density plots of the orbital rotation velocity, \vphi,
for stars in the B\&M sample (left panels) and the sample of RAVE stars from
\citet{ruchti2011} (right panels). The panels labeled
``Disk sample'' only include stars that lie below the line drawn in
Figure~\ref{lindblad}, and are expected to be dominated by stars in the
disk system. The black vertical lines indicate the mean rotational
velocity for each subsample, \mvphi. The upper grouping of panels shows
the subsets of stars in a metallicity regime ($-0.8 < $ [Fe/H] $\le
-0.5$) that is expected to be mainly occupied by thick-disk population
stars. The middle grouping of panels is the subset of
intermediate-metallicity ($-1.8 <$ [Fe/H] $\le -0.8$) stars, which are
expected to include stars from both the inner-halo population and the
MWTD. The lower grouping of panels gathers the stars in the Disk samples
together, in order to contrast the difference in the \vphi\
distributions between the canonical thick disk and the MWTD. Stars with
errors in any of the individual derived components of motion exceeding
50 \kms\ are excluded. }
\label{vphi}
\end{figure}

We now examine, in greater detail, the motions of the stars isolated
above by their location in the Lindblad diagrams. The top row of the
upper grouping of panels in Figure~\ref{vphi} shows stripe density plots
of the distribution of \vphi\ for the full sets of stars with
metallicities $-0.8 < $ [Fe/H] $\le -0.5$ from the B\&M (left) and
Ruchti et al. (right) samples. The lower row in this grouping shows the
stars in this same metallicity interval selected to the right of the
segregation line in Figure~\ref{lindblad}, labeled as the ``Disk
sample.'' This metallicity interval is expected to primarily comprise
thick-disk stars, but some overlap with the thin disk and MWTD is
inevitable. Stars that are likely members of the disk system are
identified by a ``1'' in the second digit of the INOUT parameter in
column (18) of Table~\ref{tab5}. Stars with metallicities in the high
range noted in this upper grouping of panels are identified by a ``1''
in the third digit of the INOUT parameter.

The middle grouping of panels are similar plots, but now for stars from
each sample with intermediate metallicities, $-1.8 < $ [Fe/H] $\le
-0.8$, which corresponds to an interval expected to have a significant
contribution of MWTD stars (again with possible overlap from other
stellar populations). Once more the top row of plots is the full sample
of stars, and the bottom row is the subset in this interval that falls
in the disk sample. Stars in the upper panel of this grouping would
presumably include members of the ``local halo'' component proposed by
\citet{morrison2009}.  Unfortunately, neither the B\&M sample nor the Ruchti et
al. sample contains a sufficient number of stars to explore this issue
in further detail.  So here we simply note the possibility of its
presence. Stars with metallicities in the intermediate range noted in
this grouping of panels are identified by a ``2'' in the third digit of
the INOUT parameter in column (18) of Table~\ref{tab5}. 

The lower grouping of panels compares the distribution of \vphi\ for the
two metallicity intervals considered previously, but only for the stars
in the disk samples. We employ a two-sample K-S test to check if the
subset of disk stars chosen from the B\&M sample and Ruchti et al.
samples in the high- and intermediate-metallicity intervals are
consistent with selection from the same parent population. Indeed, this
test cannot reject the common-parent null hypothesis for either
metallicity interval, in spite of the fact that the samples were clearly
chosen in different ways. In contrast, it is unsurprising that the same
K-S test applied {\emph{between}} the disk stars in the different
metallicity intervals for each of the B\&M and Ruchti et al. samples
clearly discriminates between their distributions of \vphi.

The black vertical lines mark the location of the \mvphi\ for each disk
subsample. The means and dispersions of the stars we would associate
with the canonical thick-disk population for the B\&M sample (the
metallicity interval $-0.8 < $ [Fe/H] $ \le -0.5$) are \mvphi\ = 216
\kms, $\sigma_{V_{\rm \phi}} = $ 37 \kms, while those for the Ruchti 
et al. sample are \mvphi\ = 229 \kms, $\sigma_{V_{\rm \phi}} = 30 $
\kms. Note that the values we obtain for the mean rotational velocity of
the thick-disk population from this crude analysis are somewhat higher,
and the velocity dispersions somewhat lower, than those reported by
\citet{carollo2010} (\mvphi\ = 182 \kms, $\sigma_{V_{\rm \phi}}$ = 51
\kms). It stands to reason that contamination from stars of the
thin-disk population in this interval may be at least in part
responsible for this result. However, it is also worth noting that
\citet{chiba2000}, and other authors since, have reported that the 
asymmetric drift of the thick disk exhibits a strong gradient with
distance above the Galactic plane, on the order of $\Delta $\mvphi\ /
$\Delta |Z| = 36$ \kms\ kpc$^{-1}$, according to Carollo et al. The B\&M
sample and the Ruchti et al. disk sample are both located very close to
the plane, and hence would be expected to be in more rapid rotation. In
the case of stars we would associate with a MWTD population (the
metallicity interval $-1.8 < $ [Fe/H] $ \le -0.8$), we obtain
\mvphi\ = 181 \kms, $\sigma_{V_{\rm \phi}} = 53 $ \kms, while those for
the Ruchti et al. sample are \mvphi\ = 166 \kms, $\sigma_{V_{\rm \phi}}
= 47 $ \kms. The rotational velocity of the MWTD in the present analysis
is again somewhat higher than reported by Carollo et al., who obtained a
mean rotational velocity of this component of \mvphi\ $\sim 100-150$
\kms, with a dispersion in the range 35 to 45 \kms, which is a little
lower than our derived value. Contamination of the B\&M and Ruchti et
al. samples from metal-poor stars of the inner-halo population is
possibly responsible for this result. The difference in the mean
rotational velocity might also be accounted for if, as speculated by
Carollo et al., the MWTD also exhibits a gradient in its rotational
velocity with distance from the plane; future tests with larger samples
should prove illuminating. 

\subsection{Distribution of [C/Fe] for the B\&M sample}

\begin{figure}[!ht]
\epsscale{1.20}
\plotone{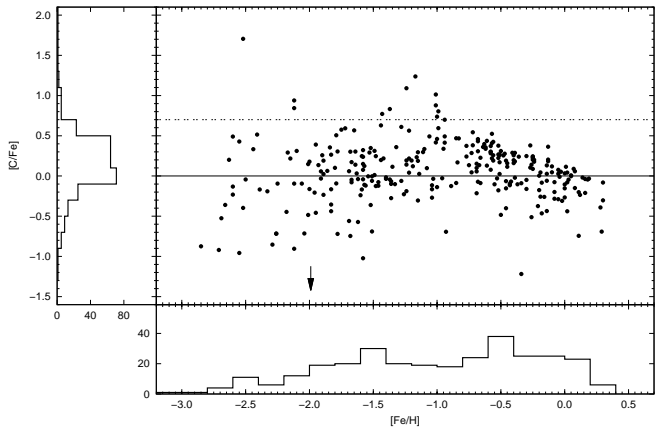}
\caption{Carbonicity, [C/Fe], as a function of the metallicity, [Fe/H], for stars
in the B\&M sample with available measurements; downward arrows indicate
derived upper limits for [C/Fe]. The marginal distributions of each
variable are shown as histograms. The horizontal dashed line marks
the definition of CEMP stars used in this work, [C/Fe] $\ge +0.7$. }
\label{cfe_feh}
\end{figure}

Figure~\ref{cfe_feh} shows the distribution of carbonicity, [C/Fe], as a
function of [Fe/H], for the stars in the B\&M sample. The general
increase in the level of [C/Fe] with decreasing [Fe/H], as has been seen
in numerous previous samples, is evident. There are 12 stars in this
sample which are classified as CEMP stars (and tagged as such in
Table~\ref{tab3}). Their frequency appears to increase with declining
[Fe/H], again as noted in previous samples. Based on their
metallicities, we expect that the majority of these stars will be
classified as CEMP-$s$, rather than CEMP-no, once high-resolution
spectroscopy has been carried out. 

\section{Discussion and Conclusions}
\label{final}

As noted in the introduction, a large number of recent papers on the
nature, origin, and evolution of the disk (and halo) system of the Milky
Way have raised new and important questions concerning issues that were
once considered ``solved.'' In this paper, we have sought to explore
what can be learned from a modern analysis of a pioneering dataset, the
``weak-metal'' sample of Bidelman \& MacConnell (1973) studied by
Norris, Bessell, \& Pickles (1985), Paper I of this series. Many of the
ideas and insights from Paper I have spawned research directions that
have not yet, even today, been completely explored to their conclusion.
Here, we have focused on resolving one of the most important claims from
that paper, the suggested existence for what has come to be known as the
metal-weak thick disk. As we have shown, the worries raised by a number
of authors, subsequent to the publication of Paper I, were indeed valid.
The faulty calibration of a photometric DDO-based metallicity
determination led to the production of a false signature in the analysis
of Paper I, which may (or may not) have unduly influenced claims for the
existence of a MWTD.

The combination of high-quality, high-S/N medium-resolution
spectroscopic data with a set of tools capable of producing accurate,
and well-tested, estimates of the atmospheric parameters (and
carbonicity, [C/Fe]) for the same sample of stars as analysed in Paper
I, has allowed us to carry out a new (and expanded) consideration of the
presence of stars that might be associated with a MWTD. We also have the
advantage that we could make use of much-improved radial velocities and
proper motions than were available to the previous study, adding
substantially to the size of the dataset suitable for kinematic study,
while simultaneously improving the quality of the derived kinematics. 

We conclude that the dataset from Paper I does indeed comprise stars
that can be associated with a MWTD. A comparison of these data with a
similar-size sample from the RAVE survey (and with atmospheric
parameters determined from a completely different set of techniques)
yields essentially the same conclusion.  We note, however, as pointed out
by an anonymous referee, that both samples of stars we have considered
are impacted by metallicity-selection biases (and in the case of the
Ruchti et al. sample, by kinematic-selection bias as well). Thus, the
relative numbers of stars present in any given metallicity interval
should clearly not be taken as representative of the underlying parent
population. We have also not addressed in this paper whether or not a
MWTD is indeed best considered a separate component from the rest of the
disk system, or whether it is somehow causually linked to the same
formation processes (themselves still actively debated) involved with
the origin of the thick (and even thin) disks. We leave these issues to
future work.

\acknowledgments 

We thank an anonymous referee for remarks which served to improve our
manuscript.  T.C.B. acknowledges partial support from grant PHY 08-22648; Physics
Frontier Center/JINA, awarded by the US National Science Foundation.
J.E.N. acknowledges support from Australian Research Council grants
DP0663562 and DP0984924. V.M.P. acknowledges support from the Gemini
Observatory. S.R. acknowledge partial support from FAPESP, CNPq, and
Capes. Y.S.L. is a Tombaugh Fellow.

\clearpage
\LongTables

\input{tab1.tex}
\clearpage

\begin{landscape}
\voffset=-1.5cm
\input{tab2.tex}
\clearpage
\end{landscape}

\input{tab3.tex}
\clearpage

\begin{landscape}
\input{tab4.tex}
\clearpage
\end{landscape}

\begin{landscape}
\input{tab5.tex}
\clearpage
\end{landscape}

\end{document}

%% file: tab1.tex
%\documentclass[preprint]{aastex}

%\hoffset=-0.5in

%\begin{document}

\begin{deluxetable}{clrrrcrcccc}
\tabletypesize{\footnotesize}
\tablenum{1}
\tablecolumns{11}
\tablewidth{0pc}
\tablecaption{Photometric Information and Adopted Reddening \label{tab1}}
\tablehead{
\colhead{BM Name} &
\colhead{Star Name} &
\colhead{LON} &
\colhead{LAT} &
\colhead{$V$} &
\colhead{$B-V$} &
\colhead{$J$} &
\colhead{$J-K$} &
\colhead{$E(B-V)_S$} &
\colhead{$E(B-V)_A$} &
\colhead{$E(B-V)_F$} \\
\colhead{ } &
\colhead{ } &
\colhead{($^{\circ}$)} &
\colhead{($^{\circ}$)} &
\colhead{ (mag)} &
\colhead{ (mag)} &
\colhead{ (mag)} &
\colhead{ (mag)} &
\colhead{ (mag)} &
\colhead{ (mag)} &
\colhead{ (mag)} \\
\colhead{ (1) } &
\colhead{ (2) } &
\colhead{ (3) } &
\colhead{ (4) } &
\colhead{ (5) } &
\colhead{ (6) } &
\colhead{ (7) } &
\colhead{ (8) } &
\colhead{ (9) } &
\colhead{ (10) } &
\colhead{ (11) } 
 }
\startdata
BM-001     & HD~702         &  94.3 & $-$69.0 &  9.81   &  0.51   &  8.767   &  0.308   &    0.034 &  0.03 &  0.02 \\
BM-002     & HD~2796        &  97.6 & $-$78.7 &  8.49   &  0.74   &  6.853   &  0.597   &    0.020 &  0.02 &  0.01 \\
BM-003     & HD~3567        & 113.1 & $-$70.9 &  9.26   &  0.46   &  8.218   &  0.329   &    0.036 &  0.04 &  0.02 \\
BM-004     & HD~3715        & 110.5 & $-$76.9 &  9.46   &  0.45   &  8.523   &  0.254   &    0.020 &  0.02 &  0.01 \\
BM-005     & HD~4306        & 118.1 & $-$72.4 &  9.01   &  0.74   &  7.424   &  0.601   &    0.036 &  0.04 &  0.04 \\
BM-006     & HD~6461        & 136.6 & $-$75.4 &  7.64   &  0.79   &  6.149   &  0.562   &    0.025 &  0.03 &  0.01 \\
BM-007     & HD~7041        & 297.8 & $-$60.6 &  9.02   &  0.80   &  7.459   &  0.558   &    0.021 &  0.02 &  0.01 \\
BM-008     & BD-11:220      & 138.6 & $-$72.7 &  9.19   &  0.58   &  8.083   &  0.365   &    0.030 &  0.03 &  0.01 \\
\enddata

\end{deluxetable}
%\end{document}

%% file: tab2.tex
%\documentclass[preprint]{aastex}

%\voffset=0.05in
%\voffset=0.75in
%\rotate
\pagestyle{empty}
%\begin{document}

\begin{deluxetable}{ccrrrrrrrccccccccccl}
\tabletypesize{\scriptsize}
\tablenum{2}
\tablecolumns{20}
\tablewidth{0pc}
\tablecaption{Radial Velocities, Line Indices, Atmospheric Parameters, and Type Assignments \label{tab2}}
\tablehead{
\colhead{BM Name} &
\colhead{N} &
\colhead{RV$_{M}$} &
\colhead{RV$_{H}$} &
\colhead{ KP } &
\colhead{ HP2 } &
\colhead{ GP } &
\colhead{ HG2 } &
\colhead{ LACF} &
\colhead{ Teff$_S$} &
\colhead{ logg$_S$} &
\colhead{ [Fe/H]$_S$ } &
\colhead{ Teff$_H$} &
\colhead{ logg$_H$} &
\colhead{ [Fe/H]$_H$ } &
\colhead{ Teff$_C$} &
\colhead{ logg$_C$} &
\colhead{ [Fe/H]$_C$ } &
\colhead{ [Fe/H]$_N$ } &
\colhead{ TYPE$_{N/S}$} \\
\colhead{ } &
\colhead{ } &
\colhead{km~s$^{-1}$ } &
\colhead{km~s$^{-1}$ } &
\colhead{ ({\AA}) } &
\colhead{ ({\AA}) } &
\colhead{ ({\AA}) } &
\colhead{ ({\AA}) } &
\colhead{ } &
\colhead{ (K)}&
\colhead{ (cgs)} &
\colhead{  } &
\colhead{ (K)} &
\colhead{ (cgs)} &
\colhead{  } &
\colhead{ (K)} &
\colhead{ (cgs)} &
\colhead{  } &
\colhead{  } &
\colhead{  } \\
\colhead{(1)} &
\colhead{(2)} &
\colhead{(3)} &
\colhead{(4)} &
\colhead{(5)} &
\colhead{(6)} &
\colhead{(7)} &
\colhead{(8)} &
\colhead{(9)} &
\colhead{(10)} &
\colhead{(11)} &
\colhead{(12)} &
\colhead{(13)} &
\colhead{(14)} &
\colhead{(15)} &
\colhead{(16)} &
\colhead{(17)} &
\colhead{(18)} &
\colhead{(19)} &
\colhead{(20)} 
}
\startdata
%STAR        N       VEL      HI      KP        HP2       GP       HG2      LACF      TEFFS   LOGGS      FEHS        TEFFH      LOGGH     FEHH      TEFFC   LOGGC      FEHC          FEHN      TYPE N/S
% (1)       (2)      (3)      (4)     (5)       (6)       (7)       (8)       (9)      (10)    (11)      (12)         (13)      (14)      (15)       (16)    (17)      (18)          (19)       (20)

BM-001  &   1  &     45 & \nodata &  7.90 &    2.85 &    3.39 &    2.53 &    0.204 &   6007  & 3.75    & $-$0.80   & \nodata & \nodata & \nodata &   6001  & 4.11    & $-$0.56   &  $-$0.44  & D   / D     \\
BM-002  &   2  &  $-$48 &   $-$61 &  6.37 &    1.06 &    1.66 &    1.25 & $-$0.347 &   5065  & 1.15    & $-$2.46   &   4901  &  1.44   & $-$2.30 &   4894  & 0.90    & $-$2.60   &  $-$2.43  & RG  / G     \\
BM-003  &   1  &  $-$32 &   $-$48 &  6.39 &    3.51 &    1.64 &    3.45 & $-$0.233 &   6105  & 3.58    & $-$1.27   &   6004  &  4.06   & $-$1.21 &   6117  & 3.90    & $-$1.14   &  $-$0.82  & D   / TO    \\
BM-004  &   1  &     29 & \nodata &  6.70 &    4.19 &    2.24 &    4.06 &    0.027 &   6355  & 3.75    & $-$0.66   & \nodata & \nodata & \nodata &   6411  & 4.11    & $-$0.39   &  $-$0.42  & D   / D     \\
BM-005  &   1  &  $-$68 &   $-$66 &  4.86 &    1.62 &    2.60 &    1.14 & $-$0.662 &   5266  & 1.70    & $-$2.42   &   4935  &  1.98   & $-$2.73 &   5130  & 1.58    & $-$2.55   &  $-$2.45  & RG  / G     \\
BM-006  &   3  &      9 &     8   & 10.01 &    0.84 &    5.65 &    1.15 &    0.472 &   5123  & 2.39    & $-$1.19   &   5116  &  2.30   & $-$0.93 &   4962  & 2.43    & $-$1.04   &  $-$1.44  & RG  / G     \\
BM-007  &   1  &    122 &   119   &  9.84 &    0.71 &    5.47 &    0.87 &    0.605 &   5285  & 2.69    & $-$0.69   &   5079  &  2.79   & $-$0.83 &   5152  & 2.80    & $-$0.42   &  $-$1.55  & RG  / G     \\
\enddata

\tablecomments{A ":" appended to column (20) indicates a disagreement between the type assignment
(based on photometry) in Paper I and that assigned based on the atmospheric parameters determined by the n-SSPP.} 
%based on the n-SSPP 
\end{deluxetable}
%\end{document}

%% file: tab3.tex
%\documentclass[preprint]{aastex}

%\hoffset=-0.5in

%\begin{document}

\begin{deluxetable}{cccccc}
\tabletypesize{\footnotesize}
\tablenum{3}
\tablecolumns{6}
\tablewidth{0pc}
\tablecaption{Carbon Abundance Ratio Estimates \label{tab3}}
\tablehead{
\colhead{BM Name} &
\colhead{[C/Fe]$_S$} &
\colhead{DETECT} &
\colhead{CC} &
\colhead{[C/Fe]$_H$} &
\colhead{[C/Fe]$_C$} \\
\colhead{ (1) } &
\colhead{ (2) } &
\colhead{ (3) } &
\colhead{ (4) } &
\colhead{ (5) } &
\colhead{ (6) } 
}
\startdata
BM-001 & $+$0.39 & D &  0.999  & \nodata & $+$0.19 \\
BM-002 & $+$0.09 & D &  0.998  & $-$0.40 & $-$0.13 \\
BM-003 & $+$0.23 & D &  0.989  & \nodata & $+$0.03 \\
BM-004 & $+$0.48 & D &  0.995  & \nodata & $+$0.29 \\
BM-005 & $+$0.61 & D &  0.994  & $+$0.26 & $+$0.43 \\
BM-006 & $+$0.05 & D &  1.000  & \nodata & $-$0.17 \\
BM-007 & $+$0.14 & D &  0.998  & $-$0.15 & $-$0.07 \\
BM-008 & $+$0.07 & D &  0.997  & \nodata & $-$0.15 \\
\enddata
\tablenotetext{a}{CEMP star ([C/Fe]$_C \ge +0.7$, DETECT = ``D'',
and CC $\ge 0.7$)}

\end{deluxetable}
%\end{document}

%% file: tab4.tex
%\documentclass[preprint]{aastex}

%\rotate
\pagestyle{empty}
%\begin{document}

\begin{deluxetable}{crrrrcrrrrrrc}
\tabletypesize{\scriptsize}
\tablenum{4}
\tablecolumns{13}
\tablewidth{0pc}
\tablecaption{Parallaxes, Distance Estimates, and Proper Motions \label{tab4}}
\tablehead{
\colhead{BM Name} &
\colhead{$\pi_{\rm HIP}$} &
\colhead{$\sigma_{\pi_{\rm HIP}}$ } &
\colhead{$D_{\rm pho}$ }&
\colhead{$D_{\rm HIP}$ } &
\colhead{$\sigma_{\pi_{\rm HIP}}/{\pi_{\rm HIP}}$} &
\colhead{$D_{\rm ado}$ } &
\colhead{$\sigma_{{\rm D}_{\rm ado}}$ } &
\colhead{$\mu_{\alpha}$ } &
\colhead{$\mu_{\delta}$ } &
\colhead{$\sigma_{\mu_{\alpha}}$ } &
\colhead{$\sigma_{\mu_{\delta}}$ } &
\colhead{HIP/TYCHO ID} \\ 
\colhead{ } &
\colhead{(mas)} &
\colhead{(mas)} &
\colhead{(kpc)} &
\colhead{(kpc)} &
\colhead{ } &
\colhead{(kpc)} &
\colhead{(kpc)} &
\colhead{(mas yr$^{-1}$)} &
\colhead{(mas yr$^{-1}$)} &
\colhead{(mas yr$^{-1}$)} &
\colhead{(mas yr$^{-1}$)} &
\colhead{ } \\ 
\colhead{(1) } &
\colhead{(2) } &
\colhead{(3) } &
\colhead{(4) } &
\colhead{(5) } &
\colhead{(6) } &
\colhead{(7) } &
\colhead{(8) } &
\colhead{(9) } &
\colhead{(10)} &
\colhead{(11)} &
\colhead{(12)} &
\colhead{(13)}
}
\startdata
% star                 plx      plxe       dpho       dhip     hiprat     dadopt err_dadopt     pmra         pmdec     pmrae     pmdece    hip/tyc
%      (1)             (2)       (3)        (4)       (5)        (6)        (7)      (8)         (9)        (10)      (11)       (12)       (13)
BM-001          &   $ 6.71$ & $ 1.40$ &   0.097  & $ 0.149$ & $ 0.21$ &   0.097 &   0.015 & $   8.82$ &  $  23.43$ &   1.50  &   0.80  &     921        \\
BM-002          &   $ 0.88$ & $ 0.81$ &   0.530  & $ 1.136$ & $ 0.92$ &   0.530 &   0.080 & $  -1.25$ &  $ -51.79$ &   0.90  &   0.59  &    2463        \\
BM-003          &   $ 9.51$ & $ 1.11$ &   0.096  & $ 0.105$ & $ 0.12$ &   0.105 &   0.012 & $  20.94$ &  $-546.76$ &   1.31  &   0.79  &    3026        \\
BM-004          &   $ 4.24$ & $ 1.14$ &   0.133  & $ 0.236$ & $ 0.27$ &   0.133 &   0.020 & $  61.47$ &  $ -22.35$ &   1.23  &   0.69  &    3134        \\
BM-005          &   $ 5.01$ & $ 1.32$ &   0.547  & $ 0.200$ & $ 0.26$ &   0.547 &   0.082 & $  60.61$ &  $  21.00$ &   1.35  &   0.94  &    3554        \\
BM-006          &   $ 3.19$ & $ 0.79$ &   0.115  & $ 0.314$ & $ 0.25$ &   0.115 &   0.017 & $  63.61$ &  $  49.81$ &   0.92  &   0.50  &    5104        \\
BM-007          &   $ 7.23$ & $ 0.86$ &   0.133  & $ 0.138$ & $ 0.12$ &   0.138 &   0.016 & $ 134.10$ &  $  12.53$ &   0.70  &   0.97  &    5455        \\
BM-008          &   $10.56$ & $ 1.01$ &   0.072  & $ 0.095$ & $ 0.10$ &   0.095 &   0.009 & $ 284.85$ &  $  -8.13$ &   0.93  &   0.72  &    5489        \\
\enddata

\end{deluxetable}
%\end{document}

%% file: tab5.tex
%\documentclass[preprint]{aastex}

%\rotate
\pagestyle{empty}
%\begin{document}

\begin{deluxetable}{crrrrrrrrrrrrrrc}
\tablewidth{0pt} \tabletypesize{\scriptsize} \tablenum{5} \tablecolumns{16} \tablecaption{Space Motions and Orbital Parameters \label{tab5}}
\tablehead{
\colhead{BM Name} &
\colhead{$R$ } &
\colhead{$Z$ } &
\colhead{$U$ } &
\colhead{$V$ } &
\colhead{$W$ } &
\colhead{$\sigma(U)$ } &
\colhead{$\sigma(V)$ } &
\colhead{$\sigma(W)$ } &
\colhead{$V_{\rm R}$} &
\colhead{$V_{\phi}$} &
\colhead{$r_{\rm peri}$} &
\colhead{$r_{\rm apo}$} &
\colhead{$Z_{\rm max}$} &
\colhead{$e$} &
\colhead{INOUT} \\ 
\colhead{ } &
\colhead{(kpc)} &
\colhead{(kpc)} &
\colhead{(km~s$^{-1}$)} &
\colhead{(km~s$^{-1}$)} &
\colhead{(km~s$^{-1}$)} &
\colhead{(km~s$^{-1}$)} &
\colhead{(km~s$^{-1}$)} &
\colhead{(km~s$^{-1}$)} &
\colhead{(km~s$^{-1}$)} &
\colhead{(km~s$^{-1}$)} &
\colhead{(kpc)} &
\colhead{(kpc)} &
\colhead{(kpc)} &
\colhead{ }     &
\colhead{ } \\
\colhead{(1)} &
\colhead{(2)} &
\colhead{(3)} &
\colhead{(4)} &
\colhead{(5)} &
\colhead{(6)} &
\colhead{(7)} &
\colhead{(8)} &
\colhead{(9)} &
\colhead{(10)} &
\colhead{(11)} &
\colhead{(12)} &
\colhead{(13)} &
\colhead{(14)} &  
\colhead{(15)} &  
\colhead{(16)}   
}
\startdata
%     star   R         Z          U        V       W     SU   SV    SW      VR      VPHI    rmin   rmax     Zmax    ecc  INOUT
%      (1)  (2)       (3)        (4)      (5)     (6)    (7)  (8)   (9)    (10)     (11)    (12)    (13)    (14)   (15)    (16)
BM-001  &  8.503 & $-0.091$ & $   1$ & $  34$ & $ -32$ &  2 &   2 &  5 & $   2$ & $ 254$ &  8.50 & 11.41 &  0.55 & 0.15 &   1/1/1  \\
BM-002  &  8.514 & $-0.520$ & $ -80$ & $-107$ & $  43$ & 11 &  16 &  4 & $ -78$ & $ 113$ &  3.05 &  9.35 &  1.01 & 0.51 &   1/1/0  \\
BM-003  &  8.514 & $-0.099$ & $-144$ & $-225$ & $ -36$ & 15 &  25 & 10 & $-144$ & $  -4$ &  0.10 & 10.54 &  0.69 & 0.98 &   1/0/0  \\
BM-004  &  8.511 & $-0.130$ & $  19$ & $ -13$ & $ -26$ &  4 &   5 &  5 & $  19$ & $ 206$ &  7.30 &  8.74 &  0.37 & 0.09 &   1/1/0  \\
BM-005  &  8.579 & $-0.521$ & $ 143$ & $ -45$ & $  82$ & 24 &   7 &  2 & $ 145$ & $ 171$ &  4.88 & 13.03 &  3.33 & 0.46 &   1/0/0  \\
BM-006  &  8.521 & $-0.111$ & $  36$ & $  15$ & $   7$ &  6 &   1 &  1 & $  36$ & $ 235$ &  7.98 & 10.29 &  0.17 & 0.13 &   1/1/2  \\
BM-007  &  8.469 & $-0.120$ & $  40$ & $ -84$ & $ -93$ &  9 &   5 &  1 & $  38$ & $ 136$ &  4.44 &  8.81 &  2.40 & 0.33 &   1/1/0  \\
BM-008  &  8.521 & $-0.091$ & $ 104$ & $ -55$ & $ -29$ & 10 &   7 &  5 & $ 104$ & $ 164$ &  4.53 & 10.45 &  0.46 & 0.40 &   1/1/0  \\
\enddata
\tablecomments{INOUT indicates membership in various subsamples
considered in this paper (see text).  The first digit takes on a value of ``1'' if
the star is accepted for the kinematic analysis, ``0'' if not.  The second
digit takes on a value of ``1'' if the star is likely to be a member of
the disk system, according to the Lindblad diagram shown in
Figure~12, ``0'' if not.  The third digit takes on a value
of ``1'' if the star is a likely disk member with metallicity in the
range $-0.8 < $ [Fe/H] $ \le -0.5$, ``2'' if it is a likely disk member in
the metallicity range $-1.8 < $ [Fe/H] $\le -0.8$, ``0'' if not.}

\end{deluxetable}

%\end{document}